\documentclass[11pt]{iopart}
\usepackage{epsf}
\usepackage{iopams}
\usepackage{graphicx}
\usepackage{array}
\usepackage{amssymb}
\usepackage{setstack} 
\usepackage{bm}
\usepackage{color}
\usepackage{mathrsfs}

\newcommand{\rhs}[1]{#1_{\mathrm{RHS}}}
\newcommand{\lprod}[3]{\left (#1,#2\right )_{{\cal D}^{#3}}}
\newcommand{\gprod}[2]{(#1,#2)_{\Omega, h}}

\newcommand{\secref}[1]{Sec.~\ref{#1}}
\newcommand{\figref}[1]{Fig.~\ref{#1}}
\newcommand{\eqref}[1]{eq.~(\ref{#1})}

\def\f{\frac}

\def\ul{\underline}

\def\lp{l_{\rm Pl}}

\usepackage{enumerate}

\newcommand{\be}{\nopagebreak[3]\begin{equation}}
\newcommand{\ee}{\end{equation}}
\newcommand{\bfig}{\nopagebreak[3]\begin{figure}}
\newcommand{\efig}{\end{figure}}
\newcommand{\ba}{\nopagebreak[3]\begin{eqnarray}}
\newcommand{\ea}{\end{eqnarray}}

\newcommand{\bmult}{\nopagebreak[3]\begin{multline}}
\newcommand{\emult}{\end{multline}}
\newcommand{\rcr}{\rho_{\mathrm{max}}}

\begin{document}
\title[Chimera]{Chimera: A hybrid approach to numerical loop quantum cosmology}

\author{Peter Diener$^{1,2}$, Brajesh Gupt$^2$ and Parampreet Singh$^2$}

\address{
$^1$ 
Center for Computation \& Technology, Louisiana State University, Baton Rouge, LA 70803, U.S.A.
}
\address{
$^2$ 
Department of Physics \& Astronomy, Louisiana State University, Baton Rouge, LA 70803, U.S.A.
}
\eads{\mailto{diener@cct.lsu.edu}, \mailto{brajesh@phys.lsu.edu}, \mailto{psingh@phys.lsu.edu}}

\begin{abstract}
The existence of a quantum bounce in isotropic  spacetimes is a key result in loop 
quantum cosmology (LQC), which has been demonstrated to arise in all the models 
studied so far. In most of the models, the bounce has been studied using numerical 
simulations involving states which are sharply peaked and which bounce at volumes much 
larger than the Planck volume. An important issue is to confirm the existence of the 
bounce for states which have a wide spread, or which bounce closer to the Planck volume. 
Numerical simulations with such states demand large computational domains, making 
them very expensive and practically infeasible with the techniques which 
have been implemented so far. To overcome these difficulties, we present an efficient  
hybrid numerical scheme 
using the property that at the small spacetime curvature, the quantum Hamiltonian 
constraint in LQC, which is a difference equation with uniform discretization in volume, can 
be approximated by a Wheeler-DeWitt differential equation.  
By carefully choosing a hybrid spatial grid allowing the use of partial differential equations 
at large volumes, and with a simple change of geometrical coordinate, we obtain a 
surprising reduction in the
computational cost. This scheme enables us to explore regimes 
which were so far unachievable for the isotropic model in LQC. Our 
approach also promises to significantly reduce the computational cost for numerical 
simulations in anisotropic LQC using high performance computing.

\end{abstract}
 
\pacs{98.80.Qc,04.60.Pp}

\maketitle

\section{Introduction}

A key prediction of loop quantum gravity (LQG) is that classical differential geometry of 
general relativity (GR) is replaced by discrete quantum geometry at the Planck scale. Loop 
quantum cosmology (LQC) which is a quantization of homogeneous spacetimes based on 
LQG, manifests this feature in the form of a discrete quantum evolution 
equation which is non-singular \cite{as1}. Due to the quantum geometric effects, 
the energy density and curvature 
invariants are bounded\footnote{This turns out to be true for all matter which satisfies the 
weak energy condition (see \cite{ps09}).}  and the classical big bang is replaced by a 
quantum big bounce. At curvature scales far smaller than the Planck scale, the quantum 
evolution equation, a finite difference equation in volume, can be well approximated by the 
Wheeler-DeWitt (WDW) differential equation, and GR is recovered in the infra-red regime. 
The smooth classical differential geometry of classical GR turns out to be a low curvature 
approximation of the underlying quantum geometry.

The existence of a bounce in LQC was first demonstrated for the case of a spatially flat 
$(k = 0)$, homogeneous and isotropic model sourced with a massless scalar field 
\cite{aps1,aps2,aps3}. In these papers, numerical simulations are performed with an 
appropriate choice of an initial state peaked on a classical trajectory at late times.  
Such a state is then  evolved backwards in internal time towards the big bang using the 
loop quantum difference equation. In the numerical evolution, 
the LQC trajectory agrees with the classical trajectory as long 
as the spacetime curvature remains much below the Planck curvature. As the 
backward evolution continues, and the curvature increases, the quantum geometric 
effects become more prominent. As a result, the LQC trajectory starts to 
deviate from the classical trajectory, eventually resulting in a bounce which occurs when 
the energy density of the massless scalar field reaches a maximum value 
$\rcr \approx 0.41 \rho_{\mathrm{Planck}}$.
Further, for states which are semi-classical at late times in a macroscopic universe, and 
which bounce at volumes much greater than the Planck volume, numerical quantum 
evolution in LQC reveals a very interesting feature. It turns out that the quantum evolution 
for such states can be well approximated by an effective continuum description derived 
from an effective Hamiltonian which incorporates quantum geometric modifications 
\cite{josh,vt,psvt}. This feature has served as an important tool to extract various results 
on the physical implications of the bounce (see \cite{as1} for a review). 

The existence of a bounce in the spatially flat isotropic model has also been studied using 
an exactly soluble model (sLQC) \cite{acs}, which is derived using a different choice of 
lapse in comparison to the works in Refs. \cite{aps1,aps2,aps3}.\footnote{In Refs. 
\cite{aps1,aps2,aps3}, the lapse was chosen to be $N=1$, whereas in sLQC the lapse is 
chosen to be $N=V$.} Though for a different lapse, sLQC confirms various properties 
which were initially discovered numerically. In particular, one can prove that the bounce is 
a generic property of all the states in the physical Hilbert space, in the sense that the 
expectation value of the Dirac observable corresponding to volume always has a global 
minima in the Planck regime. Further, there exists a universal maxima for the energy 
density which turns out to be the same as $\rcr$, and relative fluctuations and peakedness 
properties are tightly constrained across the bounce \cite{cosmic_recall,montoya_corichi2,kaminski_pawlowski}. Recently, sLQC has been used to 
compute consistent quantum probabilities for the occurrence of a bounce, which turns out 
to be unity \cite{craigh-singh13}.

Over the past few years, the prediction of the existence of a quantum bounce  
has been extended to other isotropic models, including spatially closed 
\cite{apsv,warsaw_closed} and open models \cite{kv,szulc_open} with a massless scalar 
field, in the presence of a positive \cite{ap} and negative cosmological constant \cite{bp} 
and the $\phi^2$ inflationary potential \cite{aps4}. 
As in the $k=0$ model, the quantum Hamiltonian constraint derived  in these models is 
non-singular and the evolution equation is a difference equation with uniform discreteness 
in volume.\footnote{The fact that the difference equation turns out to be uniform in volume 
is not a coincidence, but is a result of the theory being physically consistent \cite{cs1}. This  
feature can also be argued using stability properties of the difference equations in LQC 
\cite{marie1,khanna_talk}. For a discussion of these issues, see Ref. \cite{ps12}.} The 
qualitative behavior of the bounce in all the models remain the same (see Refs. 
\cite{ps12,cartin_khanna_rev} for a review of various numerical results in different 
models).

So far, the numerical studies in different isotropic models in LQC have been 
performed by considering sharply peaked states with a Gaussian profile  with 
a large scalar field momentum. The motivation for considering states with large 
field momentum comes from the observation that such states correspond to universes 
which describe a large classical universe at late times \cite{aps3,apsv}. In  numerical 
simulations of such states, it turns out that the bounce 
takes place when the energy density is very close to its maximum value, 
and the effective dynamical trajectories agree extremely well with the LQC 
trajectories throughout the evolution. However, similar investigations for states 
which are widely spread or peaked at small values of field momentum are not 
available. Here we note that Gaussian states which correspond to small field 
momentum, typically  are also widely spread in the volume. As an example, a 
Gaussian state peaked at $p_\phi = 20 \sqrt{G}  \hbar$ with a relative dispersion 
of 10\%, has a large relative spread in volume: $\Delta V/V \approx 3.2$. Thus, 
it is very widely spread in volume. This is in contrast to an initial state peaked at 
$p_\phi = 1500 \sqrt{G}  \hbar$ with the same relative dispersion in $p_\phi$, which 
yields the relative spread in volume to be approximately $\Delta V/V \approx 0.1$. 
An important question is whether the properties of the bounce which have been 
obtained so far for sharply peaked Gaussian states with large field momentum 
hold true also for states which are not sharply peaked or have small $p_\phi$. 

The answer to the above question is important for several reasons. First, it provides 
test of the robustness of the bounce for states corresponding to universes which 
are more quantum than those considered so far. This issue is especially important 
to understand in the models which are not exactly solvable. Secondly, it serves as 
an important test for the validity of the effective dynamics in LQC in the region 
where the bounce happens much closer to the Planck volume.  Note that the 
states which have been considered so far in numerical simulations in LQC, 
are peaked on a large value of $p_\phi$. Such states bounce at volumes much larger 
than the Planck volume. As an example, if the momentum of the scalar field 
$\phi$ is chosen to be $p_\phi = 1500 \sqrt{G}  \hbar$, then the bounce occurs at 
a volume greater than 1600 times the Planck volume for a Gaussian state with a 
relative fluctuation in $p_\phi = 4\%$ in the flat isotropic model. Though this volume is 
small, it is too large to explore the extreme regime of validity of the effective 
dynamics in LQC. Finally, such a study sets the stage for a more ambitious analysis of 
non-trivial generalizations of initial states, such as squeezed states. Here we note that 
using sLQC, it has been analytically shown that the key properties of the bounce, such as 
the energy density at which bounce occurs, depend on the squeezing parameter of the 
initial state \cite{montoya_corichi2}. It is therefore pertinent to generalize the numerical 
simulations of semi-classical states in LQC to include widely spread and small $p_\phi$ 
states.  

Numerical simulations of more general states, such as those with a wide spread, pose severe numerical 
challenges which are beyond the capability of the current numerical techniques in 
LQC. The main limitation comes from the Courant-Friedrichs-Lewy (CFL) stability 
criterion which gives a condition for the numerical stability of 
finite difference methods. Since the discreteness in volume in the quantum difference equation in LQC is fixed by the 
underlying quantum geometry, the CFL stability limits the size of time steps taken during the numerical 
evolution of the state. It turns out that for a stable numerical evolution, the size of 
the time step has to be inversely proportional to the number of grid points in the 
spatial grid. This can be computationally very expensive to achieve. To give an estimate, a simulation of a sharply peaked state with large $p_\phi$, say at $p_\phi = 1500 \sqrt{G}  \hbar$ with a relative fluctuation in $p_\phi$ of around 4\%  can be successfully performed by 
choosing $\sim 30,000$ points on the volume grid. This takes approximately 4 
minutes on a 2.4 GHz Sandybridge workstation with 16 cores. In contrast, for a simulation of a 
widely spread state with $p_\phi = 20 \sqrt{G}  \hbar$ it is necessary to use about $1.5\times 10^{19}$ 
spatial grid points. This requires $1.25\times 10^{14}$ times more grid points as 
compared to the sharply peaked state. Also, due to the CFL stability conditions 
one would have to take $1.25\times 10^{14}$ times smaller time steps. Such a 
simulation would take $1\times 10^{27}\, {\rm hours}\sim 10^{23}\,{\rm years}$ on a similar 
workstation. 

These limitations are not only confined to the isotropic models in LQC. 
They become more severe in the case of 
numerical simulations in anisotropic models even for sharply peaked states. For a 
locally rotationally symmetric (LRS) Bianchi-I model with a massless scalar field, one has 
to consider a two dimensional spatial grid 
and a grid in internal time. In comparison to the isotropic model, the number of spatial 
grid points $n$ scale as $n^2$ and the computational cost roughly scales as $n^3$ 
(with one power of $n$ originating from the CFL condition). For  Bianchi-I models 
which are not locally  rotationally symmetric, the computational cost scales as $n^4$ 
in comparison to the isotropic models. Hence, anisotropic models are computationally
extremely expensive. A rigorous analysis of evolution of states with quantum evolution 
equations in Bianchi models necessarily requires going beyond the techniques currently 
used in numerical LQC.  This is important to understand the quantum analog of mixmaster behavior in Bianchi models in LQC, on the lines of investigations in classical theory \cite{bb,garfinkle}.

The goal of this paper is to investigate these numerical challenges and develop an 
efficient scheme (which we call `Chimera') to tackle them, as well as establish a more general infrastructure 
for numerical simulations in LQC. Utilizing the fact that, in the 
large volume limit, the LQC evolution equations can be very well approximated by 
the WDW equation, we propose a hybrid numerical spatial grid 
composed of two parts. The inner part corresponding to the discrete LQC grid,  while the outer part, corresponding to the large 
volume limit, is taken to be a continuum WDW grid. We solve the LQC difference equation on the 
inner grid and the WDW equation on the outer grid. Since, the 
WDW equation is a partial differential equation we can use advanced PDE 
techniques to solve the evolution equation on the outer grid. We will implement two 
such methods: 
(i) a finite difference (FD) implementation, and 
(ii) a discontinuous Galerkin (DG) implementation. 
The latter scheme 
turns out to be particularly efficient computationally. It uses a smaller
number of grid points and allows larger time steps. 
With this scheme the simulation which would have taken 
$10^{23}\,{\rm years}$ with the usual method, can now be done in a few hours 
on the same workstation. Since the DG implementation is more efficient than the FD 
implementation, we have used the DG implementation to obtain the results and 
discuss various convergence tests of Chimera scheme in this paper. 
While the computational details and robustness tests of the numerical 
scheme presented here are focused towards the isotropic LQC 
with Gaussian states, the scheme has much wider applications.
Our approach not only provides an efficient numerical infrastructure for widely 
spread Gaussian states, but also proves to be very efficient to study the numerical 
evolution of non-Gaussian states in isotropic spacetimes, and as well as 2+1 and 3+1 
dimensional anisotropic spacetimes. These additional applications of the Chimera 
scheme will be presented in upcoming papers \cite{dgms,dgms2}. In this sense, the present article 
can be considered as the first in a series of papers dedicated to numerical studies in LQC.

The paper is organized as follows. In the next section we will briefly describe the 
WDW theory and LQC of the spatially flat FRW 
spacetime with a massless scalar field, obtain the LQC difference equation and discuss the way it is approximated by the WDW equation in the large volume limit. This step is largely based on the work in Ref. \cite{aps3}, to which we refer the reader for more details. We 
will discuss the construction of initial data and the large volume limit of the LQC 
difference equation which are of key importance in the development of the Chimera 
scheme. In~\secref{sec:chimera}, we give the numerical details of the Chimera
scheme. Here we consider the large volume limit of the LQC 
evolution equation and perform the Courant stability analysis by computing the 
characteristic speeds of the evolution equations. We explain the way LQC and WDW grids can be chosen to substantially decrease computational costs.  We will also discuss a few selected results for two extreme 
cases, one sharply peaked and one widely spread state, using the Chimera scheme in \secref{sec:results}.  A more detailed 
analysis of the results of the numerical evolutions of widely spread states will be provided in an 
upcoming paper \cite{dgs2} where we will analyze in detail the spread dependent properties of 
the states and compare with the effective theory for various parameters of the initial state. 
In~\secref{sec:testchimera} we perform several convergence tests of the Chimera 
scheme to check the robustness of the results. We conclude with a summary and 
discussion in \secref{sec:discussion}.

\section{Quantization of spatially flat isotropic FRW cosmology}\label{sec:lqc}

In this section we briefly review the loop quantization of a homogeneous and isotropic
flat FRW universe. The discussion in this section is mainly based on 
\cite{aps2,aps3}. We will first describe the Wheeler-DeWitt quantization in terms of the variables commonly used in geometrodynamics, and then 
move to loop quantization where we
introduce the loop variables: the 
Ashtekar-Barbero connection and the conjugate densitized triad.  
The metric for the flat FRW spacetime is given by
\be
ds^2 = -N^2dt^2 + a^2(t)\left(dx^2+dy^2+dz^2\right),
\ee
where $N$ is the lapse function and $a(t)$ is the scale factor of the universe. We choose the lapse as $N=1$. The spatial topology 
of the above metric is taken to be $\mathbb R^3$. In order to define the Hamiltonian 
framework in non-compact models and to be able to define a symplectic structure 
on the spatial manifold we need to introduce a fiducial cell $\mathcal V$, to which 
all our integrations are restricted. The volume
of the fiducial cell is chosen to be $V_o=\int d^3x \sqrt{\mathring q}$, where 
$\mathring q$ is the determinant of the fiducial metric $\mathring{q}_{ab}$ on
the spatial manifold. 
We will consider the FRW spacetime with a massless scalar 
field as the matter source. The scalar field being massless, turns out to be 
a monotonic function of time in the classical theory which, in turn, serves as a 
viable choice of relational time in the quantum theory. We will make this choice of 
clock and define our Dirac observables: a volume observable and the field momentum $p_\phi$ with respect to it. We begin with a summary of WDW quantization of this model, which is followed by the quantization of this model in LQC.

\subsection{Wheeler-DeWitt Theory}
In the WDW theory, the canonical phase space variables are: the scale factor $a$ 
and its conjugate momentum $p_{a}(=-a\dot a)$ for the gravitational sector and the 
scalar field $\phi$ and the field momentum $p_\phi (=\dot{\phi} V_o a^3)$ for the 
matter sector. Here the dot denotes the derivative with respect to proper time `t'.
These two pairs of canonical variables satisfy the following Poisson bracket relations:
\be
\{a,\,p_{a}\}=\f{4\pi G}{3 V_o}, \qquad \{\phi,\,p_\phi\}=1 ~.
\ee
The full Hamiltonian constraint is given by,
\be
\label{eq:wdwconstraint}  C^{\rm WDW} = -\f{3}{8\pi G} \f{p_a^2\, V}{a^4} + \f{p_\phi^2}{2 V} \approx0.
\ee
Using Hamilton's equations, the Hamiltonian constraint can now be solved to obtain the 
classical solution of the scale factor in terms of the proper time t as follows:
\be
a(t) = \left(\pm \sqrt{12\pi G}\, p_\phi \left(t-t_o\right)\right)^{1/3},
\ee
where $t=t_o$ corresponds to the time at which the classical singularity occurs, 
and the plus and minus signs denote the expanding and contracting solutions 
respectively. From the solution obtained above, it is clear that the scale factor $a$ 
is a monotonic function of time $t$. That is, in forward evolution, it 
always increases in an expanding universe and decreases to zero in a 
contracting universe. 

In a similar way, we can obtain the equations of motion for the matter 
field. Solving them we get: 
\be
\phi(t) = \sqrt{\f{1}{12\pi G}} \ln\left(|t-t_o|\right) + \phi_o,
\ee
where $\phi_o$ is a constant of integration that is fixed by the 
initial conditions. It is important to note that $\phi$ is a 
monotonic function of time. These solutions can be utilized 
to express the scale factor in parametric form, with $\phi$ being the parameter, as follows
\be
a\left(\phi\right) = \exp\left(\pm \sqrt{\f{4\pi G}{3}} (\phi-\phi_o)\right),
\ee
where, as noted before, the plus and minus signs denote the expanding an 
contracting  solutions respectively. Considering the scalar field $\phi$ 
as an emergent ``clock'', it is  evident that there is a past classical 
singularity at $\phi\rightarrow -\infty$ in the expanding branch and a 
future singularity at $\phi\rightarrow \infty$ in the contracting branch. 

To construct the quantum theory, one starts with the kinematical Hilbert space ${\cal H}_{\mathrm{kin}}^{\mathrm{wdw}} = L^2(\mathbb{R}^2,d a \, d \phi)$,
on which the 
scale factor and scalar field operators acts by multiplication,  
and $p_a$ and $p_\phi$ act by differentiation. 
The WDW Hamiltonian constraint (\eqref{eq:wdwconstraint}) expressed in terms of these operators yields:
\be
\label{wdweqn}\f{\partial^2}{\partial\phi^2}\ul{\Psi}( v,\, \phi) = 12 \pi G v\f{\partial}{\partial v}\left(v\f{\partial}{\partial v}\right) \ul{\Psi}( v,\,\phi) =: - \ul{\hat \Theta} \ul{\Psi}(v,\phi),  \label{eq:wdwevol}
\ee 
where, to facilitate comparison with the quantum evolution equation in LQC, we have expressed the scale factor in terms of the variable $v$ which is related to the volume $V=V_o\,a^3$ via \eqref{eq:vdef}. \Eref{eq:wdwevol} is a wave equation, with the 
scalar field $\phi$ playing the role of time, and $\widehat{\underline\Theta}$ playing the role of a spatial Laplacian. 
The physical inner product between two 
states $\ul{\Psi}_1$ and $\ul{\Psi}_2$ 
in the WDW theory is given by:
\be
\langle \ul{\Psi}_1|\ul{\Psi}_2\rangle = \int\f{dv}{v}\, \bar{\ul{\Psi}}_1(v, \phi)\,\ul{\Psi}_2(v, \phi).
\label{eq:wdwinprod}
\ee
With respect to this inner product, the physical observables --  the Dirac 
observables of the theory, which in this case are the field momentum $p_\phi$
and volume computed at constant $\phi$ slices $v|_\phi$, have a self-adjoint action. Their actions are given by \cite{aps3,acs}
\be
\hat p_\phi \ul{\Psi}(v,\phi) = \sqrt{\ul\Theta} \ul{\Psi}(v,\phi), ~~ \mathrm{and} ~~ \hat v|_{\phi_o} \ul{\Psi}(v,\phi) = e^{i \sqrt{\ul{\Theta}} (\phi - \phi_o)} v \ul{\Psi}(v,\phi).
\ee

A general solution to~\eqref{wdweqn} can be expanded in terms of the eigenfunctions of the $\ul{\Theta}$ operator: 
\be
\ul{\hat \Theta} \ul{e}_k(v) = \omega^2 \ul{e}_k(v), ~~ \ul{e}_k(v) = \f{1}{\sqrt{2\pi}} e^{i k \ln(v)},
\ee 
where $k^2 = \omega^2/12 \pi G$, as:
\be
\ul{\Psi}_\phi\left(v,\,\phi\right) = \int_{-\infty}^{\infty} d k\, \left(\tilde{\Psi}_+\left(k\right) \ul{e}_{k}(v)\, e^{i\omega \phi} + \tilde{\Psi}_-\left(k\right) \ul{{\bar{e}}}_{k}(v)\, e^{-i\omega \phi}\right).
\label{eq:wdwsol}
\ee
Here, $\Psi_+$ and $\Psi_-$ represent the positive and negative 
frequency solutions, respectively. If $\Psi_{\pm}(k)$ has support on positive values of 
$k$, the state is referred to as ``ingoing'' or ``contracting'', 
otherwise, if $\Psi_{\pm}(k)$ has support on negative values of 
$k$, it is said to be ``outgoing'' or ``expanding''. The physical Hilbert space can be decomposed into positive and negative frequency subspaces, which are preserved by the action of the Dirac observables. In this analysis we will construct the initial state corresponding to the positive frequency. Further, we will choose the state such that it is peaked on an expanding trajectory. An example of such a state is one with a Gaussian profile: 
\be
\ul{\Psi}_\phi\left(v,\,\phi\right) = \int_{-\infty}^{\infty} d k\, e^{-\left(k-k^*\right)^2/2\sigma^2} \ul{e}_{k}(v)\, e^{i\omega \phi}.
\ee
Using the above initial state,  we can now compute the expectation values 
of the Dirac observables using the inner product given in \eqref{eq:wdwinprod}. 
It turns out that the WDW trajectory obtained in this way exactly follows the classical trajectory 
during the entire evolution. As a result, in the WDW theory, the classical singularity remains 
unresolved in the sense that expectation values of the volume observable vanish \cite{aps3,acs}. This conclusion has also been reached recently at the level of probabilities in a WDW quantum universe, where it has been shown that the probability for a WDW universe to be singular turns out to be unity \cite{craig_singh11}.

\subsection{Loop quantum cosmology of $k=0$ FRW spacetime}
We now summarize the loop quantization of the spatially flat, isotropic and homogeneous model with a massless scalar field. The fundamental variables in LQG are SU(2) 
Ashtekar-Barbero connections $A_a^i$ and densitized triads $E_i^a$. Utilizing 
the symmetries of homogeneity and isotropy, these variables can be 
written in a rather simple form as follows \cite{as1}
\be
A_a^i = c~V_o^{1/3}~ \mathring{\omega}_a^i, \quad  {\rm and} \quad E_i^a=p~V_o^{-2/3}\sqrt{\mathring q} ~\mathring{e}_i^a,
\ee
where $(\mathring{\omega}_a^i,\,\mathring{e}_i^a)$ are pairs of orthonormal cotriads and 
triads, compatible with the fiducial metric $\mathring{q}_{ab}$. The symmetry 
reduced connection $c$ and the triad $p$ satisfy the following Poisson bracket relation,
\be
 \{c,\,p\} = \f{8\pi G\gamma}{3},
\ee
where $\gamma \approx 0.2375$ is the Barbero-Immirizi parameter whose numerical 
value is fixed via the black hole entropy computation in LQG 
\cite{Meissner_gamma, dl_gamma}.\footnote{The value of $\gamma$ depends on the 
way the states are counted in the black hole entropy calculations. 
However, the qualitative features of the results 
presented in this paper are independent of the numerical value of $\gamma$.}
The Hamiltonian constraint can be written in terms of the Ashtekar variables $(A,\,E)$ and 
then expressed in terms of the symmetry reduced variables $(c,\,p)$ as follows:
\be
\label{eq:hamilt} C_{\rm g} = -\f{1}{\gamma^2} \int d^3x~\varepsilon_{ijk}~ \f{E^{ai}E^{bj}F^k_{ab}}{\sqrt{|\rm{det}(E)|}} = -\f{6}{\gamma^2}c^2~\sqrt{|p|},
\ee
where $F_{ab}^k$ denotes the  field strength, 
and as in the WDW theory, we have chosen the lapse function to be $N=1$. The modulus sign over the triad is introduced because the triad can have positive or negative orientation. Since we will not be considering any fermions in this model, the choice of the orientation should not affect the physics. In the quantum theory, we will consider states which will be symmetric under this choice.

In the quantization of the  Hamiltonian constraint, the matter part is quantized
following the standard Schr\"odinger quantization, and the gravitational part is loop quantized. 
In the loop quantization of the gravitational part of the Hamiltonian constraint, the terms in 
$C_{\rm g}$ are first expressed in terms of the elementary variables: the holonomy of the connection and the flux corresponding to the triad. Due to the homogeneity, the latter turns out to be proportional to the triad itself, and the holonomy of the connection along the $k$th straight edge $\mu ~\mathring{e}_i^a$ can be written as:
\be\label{eq:holonomy}
h_k^{(\mu)} = \cos \frac{\mu c}{2} \mathbb{I} + 2 \sin \frac{\mu c}{2} \tau_k ~.
\ee
where $\mathbb I$ is the Identity matrix and $\tau_k$ is the basis of the su(2) Lie 
algebra. A crucial input of the underlying quantum geometry on the quantization of the Hamiltonian constraint manifests itself in terms 
of the field strength operator. To construct this operator, one considers holonomies of connection over a closed square loop $\Box_{ij}$: 
$h_{\Box_{ij}} = h_i^{(\bar \mu)} h_j^{(\bar \mu)} (h_i^{(\bar \mu)})^{-1} (h_j^{(\bar \mu)})^{-1}$. This loop, with area $\bar \mu^2 |p|$ is shrunk to the minimum  eigenvalue of the area in loop quantum geometry: $\Delta = 4 \sqrt{3} \pi \gamma \lp^2$ \cite{aps3,awe2}. Thus, $\bar \mu$ is related to the triad as: $\bar \mu = \sqrt{\Delta/|p|}$. If $\mu$ was a constant, holonomies (\eqref{eq:holonomy}) would act on triad eigenstates with a simple translation. However, due to the dependence on the triad, their action is not a simple translation on the triads. It turns out that the action is simple translative on the eigenkets $|v\rangle$ of the volume operator $\hat V = \widehat{|p|^{3/2}}$ \cite{aps3}:
\be
  \widehat{\exp\left(i\bar\mu c/2\right)}|v\rangle = |v+1\rangle,
\ee
where 
\be
v = K \left(\f{6}{8\pi\gamma}\right)^{3/2} V, ~~~\mathrm{and}~~~\qquad K = \f{2}{3\sqrt{3 \sqrt{3}}} ~.
\label{eq:vdef}
\ee

In addition to understanding the action of holonomies, we also need to understand the way inverse triad terms in the Hamiltonian constraint are regulated in the quantum theory. Using Thiemann's methods \cite{Thiemann:2007zz,Thiemann:1997rt}, the inverse triad operator is 
\be
  |p|^{-1/2} = {\rm sgn}(p) \left[\f{1}{4\pi \gamma \lp^2 \bar\mu} {\rm Tr} \sum_k \tau^k h_k^{(\bar\mu)} \{h_k^{(\bar\mu)-1},\, V^{1/3}\} \right].
\ee
Using the action of various operators in the gravitational part of the Hamiltonian, 
we can write the operator $\widehat{C_{\rm g}}$ which acts on the state 
$\Psi$ in the following manner:
\be
\hskip-0.3cm \widehat{C_{\rm g}} \Psi\left(v,\,\phi\right)  =  C^+(v) \Psi\left(v+4,\,\phi\right) + C^o(v)\Psi\left(v,\,\phi\right) + C^-(v) \Psi\left(v-4,\,\phi\right)
\ee
where the coefficients $C^+,\,C^-\,{\rm and}\, C^o$ are given by \cite{aps3}:
\ba
  C^+(v) &=& \frac{3\pi K G}{8} |v+2|\left | |v+1|^{}-|v+3| \right |, \nonumber \\
  C^-(v) &=& C^+(v-4)=\frac{3\pi K G}{8} |v-2|\left | |v-3|-|v-1| \right |, \label{eq:lqccoeff}\\
  C^0(v) &=& -C^+(v)-C^-(v). \nonumber 
\ea
The full Hamiltonian constraint can then be written as
\be
\label{eq:lqcevol}  \f{\partial^2}{\partial \phi^2} \Psi(v,\,\phi) = -\widehat\Theta \Psi(v,\,\phi),
\ee
where the operator $\Theta$, a difference operator, is the evolution operator 
of LQC
\be
\hskip-1.6cm \widehat\Theta \Psi(v,\,\phi) = -{B(v)}^{-1} \left[C^+(v) \Psi\left(v+4,\,\phi\right) \right. 
   + C^0(v)\Psi\left(v,\,\phi\right) 
   \left. + C^-(v) \Psi\left(v-4,\,\phi\right)\right]
\ee
and the function $B(v)$, denoting the eigenvalues of $\widehat{|p|^{-3/2}}$, is given by
\be
  B(v) = \f{27 K}{8} |v|\arrowvert |v+1|^{1/3} - |v-1|^{1/3}\arrowvert^3.
\ee
In contrast to the WDW evolution operator, $\widehat\Theta$ is a quantum difference operator, and the resulting evolution equation is non-singular.  The physical inner product between two states $\Psi_1(v,~\phi)$ and $\Psi_2(v,~\phi)$ is given as
\be
\langle\Psi_1|\Psi_2\rangle = \sum_v B(v) \bar\Psi_1(v,~\phi_o) \Psi_2(v,~\phi_o)
\label{eq:innerprod}
\ee
at the emergent time $\phi=\phi_o$. The physical states are considered to be symmetric under the change of orientation of the triad, i.e. they satisfy: $\Psi(v,\phi) = \Psi(-v,\phi)$. Since the quantum evolution operator $\widehat \Theta$ couples a physical state on a uniform lattice in volume, the wavefunctions have support on lattices $\pm \epsilon + 4 n$, where $\epsilon \in [0,4)$. It is important to note that these lattices are preserved by quantum dynamics, and the physical Hilbert space becomes separable in $\epsilon$ sectors. In our analysis, we focus on the case of $\epsilon = 0$ which includes the possibility of vanishing volume in the quantum evolution. 
As in the WDW theory, the physical observables are  the set of 
Dirac observables:  $p_\phi$ and $v|_{\phi=\phi_o}$. These observables are self-adjoint with respect to the inner product (\eqref{eq:innerprod}).

Using the inner product defined in~\eqref{eq:innerprod} the expectation value of 
$\widehat{v}$ and $\widehat{p_\phi}$ can be computed as follows
\ba
\langle \Psi|\widehat{v}|_{\phi=\phi_o}|\Psi\rangle&=&||\Psi||^{-1} \sum_v B(v) |v||\Psi(v,~\phi_o)|^2, \nonumber \\
\langle \Psi|\widehat{v^2}|_{\phi=\phi_o}|\Psi\rangle&=&||\Psi||^{-1} \sum_v B(v) |v^2||\Psi(v,~\phi_o)|^2, \label{eq:expectation}\\
\langle \Psi|\widehat{p_\phi}|\Psi\rangle&=&||\Psi||^{-1} (-i \hbar)\sum_v B(v) \bar{\Psi}(v,~\phi)~\partial_\phi \Psi(v,~\phi), \nonumber \\
\langle \Psi|\widehat{p_\phi^2}|\Psi\rangle&=&||\Psi||^{-1} (-i \hbar)^2\sum_v B(v) \bar{\Psi}(v,~\phi)~\partial_\phi^2 \Psi(v,~\phi), \nonumber
\ea
where $||\Psi||=\langle\Psi|\Psi\rangle$. These can then be used to compute 
the dispersions in $p_{\phi}$ and $v$ as
\be
\label{eq:dispersion}\langle\Delta \widehat{p_\phi}\rangle = \sqrt{\langle\widehat{p_\phi^2}\rangle-\langle\widehat{p_\phi}\rangle^2}, \qquad \langle\Delta \widehat{v}|_{\phi=\phi_o}\rangle = \sqrt{\langle\widehat{v^2}|_{\phi=\phi_o}\rangle-\langle\widehat{v}|_{\phi=\phi_o}\rangle^2}. 
\ee
For the brevity of notation, in the discussion of results and plots, we will denote 
$\langle\Delta \widehat{V}|_{\phi=\phi_o}\rangle$ as $\Delta V$ 
(or equivalently $\langle\Delta {V}|_{\phi=\phi_o}\rangle$ as $\Delta V$) for the 
dispersion in volume, 
and similarly $\langle\widehat{V}|_{\phi=\phi_o}\rangle$ as $V$ for the volume observable.

Let us now turn to the discussion of the large volume limit and construction of initial data 
for the evolution. In the limit of large volume $v\gg 1$ we find that the LQC coefficients, given in~\eqref{eq:lqccoeff}, take the form
\begin{eqnarray}
  C^+(v) = \frac{3\pi K G}{4} (v+2), \nonumber \\
  C^-(v) = C^+(v-4)=\frac{3\pi K G}{4} (v-2), \\
  C^0(v) = -C^+(v)-C^-(v)=-\frac{3\pi K G}{2} v, \nonumber\\
  B(v) = \frac{K}{v} + \mathcal{O}\left(v^{-3}\right). \nonumber
\end{eqnarray}
Using these we find that the action of spatial operator $\widehat \Theta$ in the large volume limit can be written, to leading order, as 
\begin{eqnarray}
  \widehat{\Theta}\Psi(v,\phi) = -12\pi G v & \left [ v
   \frac{\Psi(v+4,\phi)-2\Psi(v,\phi)+\Psi(v-4,\phi)}{16}\right. \nonumber \\
 & \left. +\frac{\Psi(v+4,\phi)-\Psi(v-4,\phi)}{8}  \right ]. \label{eq:wdwfd}
\end{eqnarray}
We recognize the two terms in square brackets in~\eqref{eq:wdwfd}
as the second order accurate finite difference approximation to the second
and first partial derivative of $\Psi(v,\phi)$ with grid-spacing $\Delta v=4$. 
Thus in the large volume limit the equation of motion can be written as
\begin{equation}
\label{wdw}
\frac{\partial^2\Psi}{\partial\phi^2} = 12\pi G v \left [ v \frac{\partial^2\Psi}{\partial v^2}+\frac{\partial\Psi}{\partial v} \right ] = 12\pi G v 
\frac{\partial}{\partial v}\left ( v\frac{\partial\Psi}{\partial v}\right ),
\end{equation}
which is the WDW equation~\eqref{eq:wdwevol}. 

The initial states are constructed such that they are peaked on an expanding  
classical trajectory at very large volume. Such states can be obtained by considering solutions of the WDW
equation at late times. Here, we construct the initial data following the method of phase rotation of eigenfunctions of the WDW theory (`method-3' in \cite{aps2,aps3}).  Considering only the positive frequency solutions, the form of the initial 
 state is 
\be
\label{eq:wdwinteg}{\Psi}\left(v,\,\phi_o\right) = \int_{-\infty}^{\infty} d k\, {\Psi}\left(k\right) \ul{e}_{k}(v)\, e^{i\omega \phi_o} e^{-i\alpha}.
\ee
As compared to the solution of the WDW equation given in equation \eqref{eq:wdwsol}, the 
above integral has an additional phase factor given by $e^{-i\alpha}$, where 
$\alpha=k (\ln(|k|)-1)$ and $k=-\omega/\sqrt{12\pi G}$. This phase is introduced 
in order to match the eigenfunctions of the WDW equation with that of the
LQC evolution operator at large volume. For a more elaborate 
discussion on obtaining the phase factor, see \cite{aps3}.
Given a form of the wavepacket $\Psi(k)$, the integral in equation 
\eqref{eq:wdwinteg} can be evaluated and the initial data can be constructed. In this article 
we choose $\Psi(k)$ to be a Gaussian peaked at $k^* = - p_\phi^*/\sqrt{12 \pi G}  \hbar$,  with a spread 
$\sigma$ such that
\be
\Psi(k) = e^{-\left(k-k^*\right)^2/2\sigma^2}.
\ee 
In the numerical simulations performed here, we numerically 
integrate~\eqref{eq:wdwinteg} for such a wavepacket
to obtain the form of the initial wave-function. The initial data can then 
be evolved using the LQC difference equation. For numerical convenience we introduce $\Phi = \partial\Psi/\partial\phi$ as
an independent variable and rewrite~\eqref{eq:lqcevol} in first order in time
form 
\ba
\frac{\partial}{\partial\phi}\Psi(v,\phi) & = & \Phi(v,\phi) 
\label{eq:lqcevol1}\\
\frac{\partial}{\partial\phi}\Phi(v,\phi) & = & -\widehat\Theta\Psi(v,\phi).
\label{eq:lqcevol2}
\ea
In this paper we will use the above evolution equations to study the 
numerical evolution of semiclassical states.

\section{The Chimera scheme} \label{sec:chimera}
In this section we describe the formulation of the Chimera scheme and two different 
methods of its implementation. Solving the 
evolution equation of LQC is equivalent to solving a one dimensional wave equation. 
The initial data, at $\phi=\phi_o$, is given in the form of the wavefunction $\Psi(v,\,\phi_o)$
and its $\phi$-derivative $\Phi(v,\,\phi_o)$, which are 
functions of the label $v$. The idea is now to 
compute $\Psi$ and $\Phi$ at each value of $v$. In this 
setting, the variable $v$ plays the role of spatial coordinate and $\phi$ the
role of time, in analogy with the one 
dimensional wave equation. Here and in what follows, we will refer to $v$ as the spatial and $\phi$ as the time coordinate of this evolution system. 

The WDW equation~\eqref{eq:wdwevol} is a wave equation written in second order
form in both time and space (in the sense that it involves second order
derivatives of both space and time). For numerical analysis (and numerical 
implementation) it is convenient to rewrite the equation in first order in time and 
space form by introducing the auxiliary variables 
$\ul{\tilde{\Phi}} = \partial\ul{\Psi}/\partial \phi$
and $\ul{\tilde{\Pi}} = \partial\ul{\Psi}/\partial v$ and convert \eqref{eq:wdwevol} into the
system of equations
\begin{eqnarray}
	\frac{\partial\ul{\tilde{\Phi}}}{\partial\phi} = 12\pi G v^2\frac{\partial\ul{\tilde{\Pi}}}{\partial v}
	+12\pi G v\ul{\tilde{\Pi}}, \nonumber \\
	\frac{\partial\ul{\tilde{\Pi}}}{\partial\phi} = \frac{\partial\ul{\tilde{\Phi}}}{\partial v},
\label{eq:wdw} \\
\frac{\partial\ul{\Psi}}{\partial\phi} = \ul{\tilde{\Phi}}. \nonumber
\end{eqnarray}
The first equation comes from the original second order equation by
substituting the definitions of $\ul{\tilde{\Phi}}$ and $\ul{\tilde{\Pi}}$.
The second equation comes from taking a derivative with respect to $\phi$
of the definition of $\ul{\tilde{\Pi}}$, and interchanging $v$ and $\phi$
derivatives. The third equation is just the definition of  $\ul{\tilde{\Phi}}$.
This set of equations is equivalent to the original second order 
equation~\eqref{eq:wdwevol}.
In both the second order and first order form, $\ul{\Psi}(v,\phi_0)$ and 
$\partial\ul{\Psi}/\partial \phi |_{\phi_0}=\ul{\tilde{\Phi}}(v,\phi_0)$ is 
needed initially in order to evolve the system with respect to $\phi$.  In the
first order form $\ul{\tilde{\Pi}}(v,\phi_0)$ is in principle needed as well,
but it can be found from $\ul{\Psi}(v,\phi_0)$ simply by taking a numerical or
analytical derivative of $\ul{\Psi}(v,\phi_0)$ with respect to $v$. 
As this set of equations does not contain any 
spatial derivatives of $\ul{\Psi}$ with respect to $v$, only the fields 
$\ul{\tilde{\Phi}}$ and $\ul{\tilde{\Pi}}$ are important for determining 
the characteristic properties.  The last equation is only
present in order to be able to obtain $\ul{\Psi}$ by integrating 
$\ul{\tilde{\Phi}}$ with respect to $\phi$.  So, in the following 
discussion we omit this equation and write the first two equations 
in~\eqref{eq:wdw} in matrix form by introducing 
${\mathbf{\tilde{u}}} = (\ul{\tilde{\Phi}},\ul{\tilde{\Pi}})$ 
as 
\begin{equation}
  \frac{\partial{\mathbf{\tilde{u}}}}{\partial\phi} = 
  {\mathbf{\tilde{A}}}\frac{\partial{\mathbf{\tilde{u}}}}{\partial v}+{\mathbf{B \tilde{u}}},
\end{equation}
where
\begin{equation}
  {\mathbf{\tilde{A}}} = \left (\begin{array}{cc}
                          0 & 12\pi G v^2 \\
                          1 & 0
                        \end{array} \right )
\end{equation}
and 
\begin{equation}
  {\mathbf{B}} = \left (\begin{array}{cc}
                          0 & 12\pi G v \\
                          0 & 0
                        \end{array} \right ).
\end{equation}
The eigenvalues of $\mathbf{\tilde{A}}$ gives the characteristic speeds of the 
Wheeler-DeWitt equation\footnote{The matrix $\mathbf{B}$ has no influence on
the characteristic properties of the system as it does not multiply the
spatial derivative of $\mathbf{\tilde{u}}$.}. We find
\begin{equation}
	\tilde{\lambda}_{\pm} = \pm \sqrt{12\pi G} v,
\end{equation}
i.e.\ the characteristic speeds are proportional to $v$. Using an explicit
time integration scheme, we are limited in time step size $\Delta\phi$ by the 
Courant-Friedrich-Lewy (CFL) condition requiring  that
$\Delta\phi < C\Delta v/\max(|\tilde{\lambda}_{\pm}|)$ for numerical stability. 
Here $C$ is a constant depending on the numerical scheme, $\Delta v$ is
the spatial grid spacing and $\max(|\tilde{\lambda}_{\pm}|)$ is the maximum 
characteristic speed on the whole 
numerical grid. Since $\max(|\tilde{\lambda}_{\pm}|)$ increases linearly with the
location of the outer boundary on the grid, this means that the maximal time
step size is inversely proportional to the outer boundary location
$v_{\mathrm{outer}}$ according to the inequality 
\begin{equation}
  \Delta\phi \le C\frac{2}{\sqrt{3\pi G}\, v_{\mathrm{outer}}}.
\end{equation}
Since the number of grid points as well as the number of time steps increases 
linearly with $v_{\mathrm{outer}}$, we find that the computational cost scales 
proportional to $v_{\mathrm{outer}}^2$. The same must then be true for the 
original LQC difference evolution equation~\eqref{eq:lqcevol2} since it is well 
approximated by the WDW equation when the outer boundary is located at a large 
enough volume.

Changing the coordinate used in the WDW equation from $v$ to $x = \ln v$
results in the following form of the system of evolution equations
\begin{eqnarray}
	\frac{\partial\ul{\Psi}}{\partial\phi} = \ul{\Phi}, \label{eq:wdw1} \\
\frac{\partial\ul{\Phi}}{\partial\phi} = 12\pi G \frac{\partial\ul{\Pi}}{\partial x}, \label{eq:wdw2} \\
\frac{\partial\ul{\Pi}}{\partial\phi} = \frac{\partial\ul{\Phi}}{\partial x},
\label{eq:wdw3}
\end{eqnarray}
where 
\begin{eqnarray}
\ul{\Phi} = \frac{\partial\ul{\Psi}}{\partial\phi} = \ul{\tilde{\Phi}}, \\
\ul{\Pi} = \frac{\partial\ul{\Psi}}{\partial x} = v\ul{\tilde{\Pi}}.
\end{eqnarray}
Here, the variables without a tilde use $x=\ln v$ as the coordinate while
the variables with a tilde use $v$ as the coordinate.
As before we can introduce ${\mathbf{u}} = (\ul{\Phi},\ul{\Pi})$ and write the equations in matrix 
form as
\begin{equation}
  \frac{\partial{\mathbf{u}}}{\partial\phi} = 
  {\mathbf{A}}\frac{\partial{\mathbf{u}}}{\partial x}, 
\end{equation}
where
\begin{equation}
  {\mathbf{A}} = \left (\begin{array}{cc}
                          0 & 12\pi G \\
                          1 & 0
                        \end{array} \right ).
\end{equation}
In this case the matrix ${\mathbf{B}}$ is zero. 
Now the characteristic speeds (the eigenvalues of $\mathbf{A}$) are 
\begin{equation}
  \lambda^{\pm} = \pm \sqrt{12\pi G},
\end{equation}
i.e.\ independent of the spatial coordinate $x$. 

The simple idea behind the Chimera hybrid scheme then is to split the
numerical domain into two separate pieces with an interface at 
$v=v_{\mathrm{interface}}$ and solve the full LQC difference 
equations~\eqref{eq:lqcevol1} and~\eqref{eq:lqcevol2} on the inner grid with fixed grid 
spacing $\Delta v=4$ and the WDW equations~\eqref{eq:wdw1}, \eqref{eq:wdw2} 
and~\eqref{eq:wdw3} using $x=\ln v$ as variable on the outer grid. Of course 
$v=v_{\mathrm{interface}}$ has to be chosen large enough that the WDW equation 
is in fact a good approximation to the full LQC difference equation. The main 
advantages to this approach are: 1) a significant reduction in the number of grid 
points since $x=\ln v$, and 2) a significant increase in the time step size, since now 
the time step is determined by the value of $v_{\mathrm{interface}}$ instead of 
$v_{\mathrm{outer}}$.

In the following we will be using the Method of Lines (MoL) where the
computational domain is discretized in the spatial (volume) direction only,
while leaving the $\phi$-direction continuous. 
The discreteness of the quantum evolution in LQC grid is fixed by the quantum theory at $\Delta v=4$, while
we can choose the discretization $\Delta x$ on the WDW grid as per our requirements.
This turns the set of equations (difference equations on the LQC grid and
discretized PDE's on the WDW grid) into a system of coupled ordinary 
differential equations (ODE's) that can be evolved in the $\phi$-direction
using standard ODE solvers. There is one ODE per grid point and the coupling 
comes from the dependence of the right hand side (RHS) of the evolution 
equations on state information from neighboring grid points (on the LQC grid
through the difference equations and for the WDW grid through the discretized
approximation to the volume derivatives).

When using a finite numerical domain, the problem is not stated fully without
specifying the boundary conditions.  
In the numerical evolution of the initial state, we need to specify the
boundary
conditions at the outer boundary. Once again utilizing the fact that at
large
volume LQC difference equation can be very well approximated by WDW
equation,
we can specify the boundary conditions by analytically evaluating
$\Psi$ from the
integral in equation \eqref{eq:wdwinteg} and its derivative approximated by
$\partial_\phi \Psi=\sqrt{12\pi G}\, v\partial_v \Psi$. Using the
Chimera scheme,
however, we can choose the outer boundary so large that the state and
its derivative are smaller than the roundoff error  
and, hence, essentially zero
at the outer boundary. Therefore, in our numerical simulations, we
choose the value of the
state and its derivative to be zero at $v_{\rm outer}$. We have tested
that numerical simulations with this choice of boundary condition gives the
same numerical results as
setting the boundary conditions by evaluating $\Psi$ and its derivative
at $v_{\rm outer}$. 

We implement this scheme in two different ways. The first uses a finite
difference (FD) and the second a Discontinuous Galerkin (DG) approximation to
the derivatives on the outer WDW grid. It turns out that the DG implementation
is much more efficient than the FD implementation, because 
it uses a lot less number of grid points on the outer grid than the FD 
implementation. Therefore, we have used the DG implementation to obtain the 
results in the later sections. 
We now present, in turn, a more detailed description of these two different
implementations of the Chimera scheme.
\subsection{Finite difference implementation}
\label{sec:fdi}
\begin{figure}
\begin{center}
\includegraphics[width=0.7\textwidth]{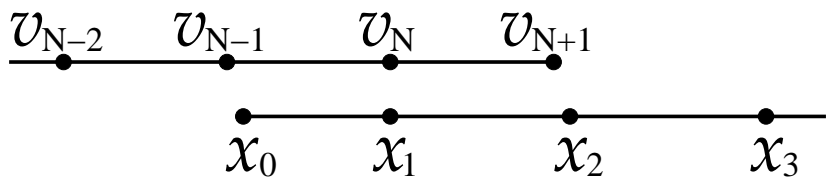}
\end{center}
\caption{The grid setup near the interface boundary in the finite difference 
implementation of the Chimera scheme. The positions of the
grid points on the outer grid are shown in $v$-space. In $x$-space they
are uniformly spaced.}\label{fig:fdgrid}
\end{figure}

We set up two separate grids with a slight overlap as indicated 
in~\figref{fig:fdgrid}.
On the inner grid, the grid spacing is uniform in $v$ with $\Delta v=4$ (which is 
fixed by the underlying discreteness of the quantum geometry) and 
$v_{\mathrm{N}}$ denotes $v_{\mathrm{interface}}$. The inner grid is extended by one
extra gridpoint ($v_{\mathrm{N}+1}$). On this grid, the grid functions 
$\Psi, \Phi$ and their ``time'' derivatives $\rhs{\Psi}, \rhs{\Phi}$ are 
allocated, where the subscript RHS denotes that the time derivatives are given via 
the right hand side of~\eqref{eq:lqcevol1} and~\eqref{eq:lqcevol2}. 
Given $\Psi$ and $\Phi$ data everywhere, $\rhs{\Psi}$ is then computed 
everywhere on the grid according to~\eqref{eq:lqcevol1} and 
$\rhs{\Phi}$ is computed up to and including the grid point $v_N$ according 
to~\eqref{eq:lqcevol2}. On the outer grid we choose uniform grid 
spacing (the simplest finite difference choice) in $x$ with 
$\Delta x=4/v_{\mathrm{interface}}$ (in order to ensure
that the resolution is the same on both sides of the interface) and $x_1 = 
\ln v_{\mathrm{interface}}$. The outer grid is extended by an additional grid
point ($x_0$) on the inside. On this grid the grid functions $\ul{\Psi},
\ul{\Phi}, \ul{\Pi}$ and their RHS $\rhs{\ul{\Psi}}, \rhs{\ul{\rho}}, 
\rhs{\ul{\Pi}}$ are allocated\footnote{The grid functions with under lines are
WDW variables while grid functions without underlines are LQC variables.}. Given $\ul{\Psi}, \ul{\Phi}$ and $\ul{\Pi}$
data everywhere $\rhs{\ul{\Psi}}$ is then computed everywhere according
to~\eqref{eq:wdw1}, while $\rhs{\ul{\Phi}}$ and $\rhs{\ul{\Pi}}$ are computed
everywhere from and including grid point $x_1$ according
to~\eqref{eq:wdw2} and~\eqref{eq:wdw3} using second order accurate centered
finite differencing. Before completing a full time step
we need to fill in values for $\rhs{\Phi}$ at point $v_{N+1}$, and for
$\rhs{\ul{\Phi}}$ and $\rhs{\ul{\Pi}}$ at point $x_0$. Since,
in the large volume limit,
$\rhs{\Phi}=\rhs{\ul{\Phi}}$ the value of 
$\rhs{\Phi}$ at point 
$v_{N+1}$ can just be filled using quadratic interpolation of the values of
$\rhs{\ul{\Phi}}$ in $x_1, x_2$ and $x_3$. Similarly the value of 
$\rhs{\ul{\Phi}}$ at point $x_0$ can be filled using quadratic 
interpolation of the values of $\rhs{\Phi}$ in $v_{N-2}, v_{N-1}$ and $v_N$.
Now, $\rhs{\ul{\Pi}}$ on the outer grid does not have a corresponding grid 
function on the inner grid. However, since 
\begin{equation}
\rhs{\ul{\Pi}}=\frac{\partial\ul{\Phi}}{\partial x} = 
v\frac{\partial\Phi}{\partial v}
\end{equation}
we can find approximations to $\rhs{\ul{\Pi}}$ in grid points $v_{N-2}, 
v_{N-1}$ and $v_N$ on the inner grid by taking second order accurate finite 
differences of $\Phi$ with respect to $v$, convert to derivatives with respect
to $x$ and interpolate those values to grid point $x_0$ on the outer grid. This 
prescription leads to a stable evolution scheme, where ingoing and outgoing
modes are passed back and forth without noticeable problems between the inner
and outer grids.

\subsection{Discontinuous Galerkin implementation}
\label{sec:gdi}
Instead of using finite difference techniques for the WDW equation on the
outer grid, we can in principle use any stable scheme for solving the PDE's.
With the second order finite difference scheme presented above, we have to
use the same resolution on either side of the interface. With a higher order
scheme we can achieve the same accuracy on the outer grid at a lower resolution,
thus reducing the computational cost even further. We chose a discontinuous
Galerkin (DG) scheme as an alternative and, since such schemes are not 
commonly used in the 
fields of numerical relativity or loop quantum gravity, we have presented the basic
ideas behind the scheme in some detail in~\ref{ap:dg}. In summary, the main
idea is to divide the computational domain into a number of elements and, within 
each element, represent the grid functions using $p$-th order polynomials, where
$p$ can be a large integer. It is possible to use either a polynomial basis
or interpolating polynomials to represent grid functions. In the latter case
$p+1$ nodes within each element are selected and the interpolating polynomials
are constructed so that the polynomial $P_i$ has the value of the grid function
at node $i$ and is zero at all other nodes. This is the approach we take in our analysis. Approximations to derivatives of grid function at all nodes can then
be computed simply by evaluating the derivative of the interpolating 
polynomials. The final ingredient is the coupling of each element to its
neighbors. This is done using suitable numerical fluxes evaluated at either
side of an element boundary. Such numerical fluxes can be constructed in 
several different ways to maintain numerical stability and accuracy. In this
implementation we use simple Lax-Friedrich fluxes (see~\ref{ap:dg}).

\begin{figure}
\begin{center}
\includegraphics[width=0.9\textwidth]{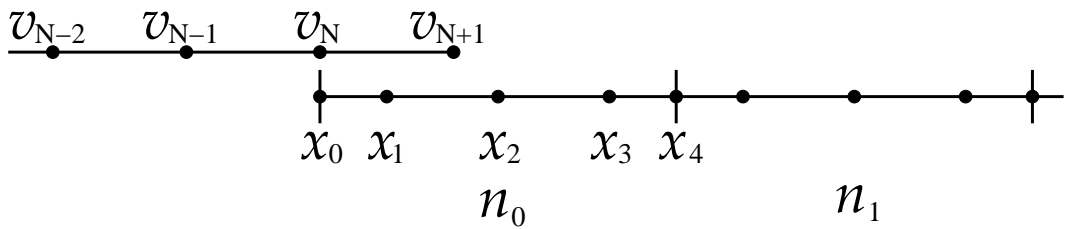}
\end{center}
\caption{The grid setup near the interface boundary in the discontinuous
Galerkin implementation is presented. It corresponds to the case of 4th order discontinuous Galerkin implementation. Here $n_0$ and $n_1$ 
label the first two DG elements. Each element consists of 5 nodes (labeled
$x_0$--$x_4$ for the first element). Node $x_4$ of the first element coincides
with the first node (unlabeled) of the second element.
}\label{fig:dginterface}
\end{figure}
The inner grid and variables are set up in exactly the same way as for the 
finite difference implementation described in~\secref{sec:fdi} and 
illustrated in~\figref{fig:dginterface}.  The outer grid is now divided 
into $n$ elements each with $p+1$ unevenly spaced nodes (where $p$ is the 
order of the scheme).  \Fref{fig:dginterface} shows the first two 
elements (labeled $n_0$ and $n_1$) for the case of a 4th order DG scheme.  
The first node in the first element 
(labeled $x_0$) is placed at $x_0 = \ln v_{\mathrm{interface}} = \ln v_N$.  
On the outer grid we use the same variables ($\ul{\Psi}, \ul{\Phi}, \ul{\Pi}$)
and their RHS as in the finite difference implementation.  The interface 
between the two grids, however, is treated differently.  In the DG scheme
we set the fields $\Psi$ and $\Phi$ in point $v_{N+1}$ on the inner grid 
directly from interpolation from the corresponding variables in the first 
element on the outer grid.  As the solution on the outer grid is stored as 
interpolating polynomials, this is a straightforward operation.  To transfer
information from the inner grid to the outer grid, we only need to construct
a numerical flux for the left boundary of the first DG element using 
information from the inner grid.  The flux needs values for $\ul{\Phi}$
and $\ul{\Pi}$ at the boundary.  These are simply given by 
\begin{eqnarray}
  \left . \ul{\Phi} \right |_{x_0} & = & 
  \left .\Phi\right |_{v_N}, \\
  \left . \ul{\Pi}\right |_{x_0} & = &
  \left . \frac{\partial\ul{\Psi}}{\partial x}\right |_{x_0} =
  \left . v\frac{\partial\Psi}{\partial v}\right |_{v_N} \approx
  v_{N} \frac{\Psi_{N+1}-\Psi_{N-1}}{8}.
\end{eqnarray}
The boundary is then treated like any other boundary between two neighboring
DG elements, independently of the order of the DG scheme.  As the accuracy
of the DG scheme is much higher than the 2nd order finite difference scheme,
we do not have to use the same resolution on the inner and outer grids.  For
most runs presented later we use 64th order DG elements with an effective
resolution 32 times lower than a corresponding 2nd order finite difference
scheme would need.  This saving is the main advantage the DG scheme has
over the 2nd order finite difference scheme.

\section{Results}\label{sec:results}
Let us now discuss a representative set of results for the simulations in the flat FRW model 
with a massless scalar field which show the implementation of the Chimera method. We will present a more detailed 
analysis of the results by considering various other values of the field momenta and 
the relative volume spreads in an upcoming article \cite{dgs2}  based on an application of the Chimera scheme. The initial data is constructed as a solution of the WDW 
equation, in the low curvature regime, by choosing the initial
state to be peaked on a large value of the volume. A large value of the volume 
ensures that the energy density of the scalar field $\rho=p_\phi^2/2\,V^2$, is 
much smaller than the Planck density. It turns out that the physical volume at the 
bounce is proportional to the value of the field momentum $p_\phi$, a constant of 
motion for the massless scalar field. Hence, a large value of the field momentum will give 
rise to a large bounce volume. 

\begin{figure}
       \includegraphics[width=\textwidth]{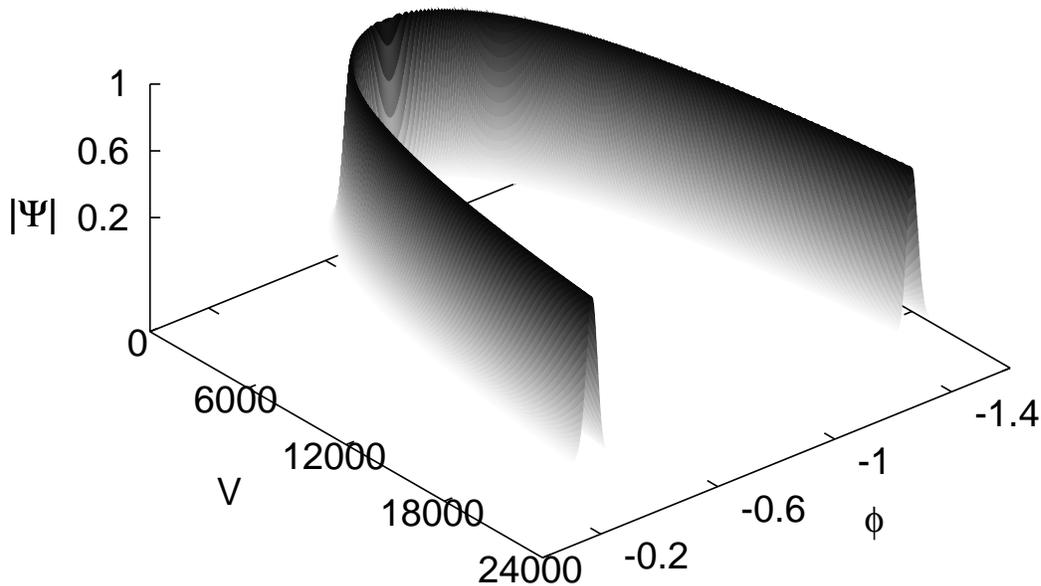}
\caption{This figure shows the time evolution of absolute value of the wavefunction. 
For clarity we only show $|\Psi|>10^{-3}$ in the plot. The values of the 
parameters are $p_\phi^*=1500\, \hbar\sqrt{G}$, $\sigma=95$ and $\Delta V/V=0.06$. It is clear 
from the figure that the state remains sharply peaked throughout the evolution.}
\label{fig:3dw1500}
\end{figure}

We will consider two extreme cases: the first corresponds to a sharply peaked state with 
a large field momentum such that the initial relative volume dispersion is very small 
compared to unity 
($p_\phi^*=1500\, \sqrt{G}\hbar,\,\sigma=95,\, \Delta V/V\approx0.06, \, V^*\approx68700\,V_{\rm Pl}$) for 
which the volume at bounce is $V_{\rm b}=1659.83\, V_{\rm Pl}$, and the second 
corresponds to a wide state with small field momentum and large initial relative volume 
dispersion ($p_\phi^*=20 \,\sqrt{G} \hbar,\,\sigma=2.25,\, \Delta V/V\approx6.44,\, V^*=68700\,V_{\rm Pl}$) for which the
volume at bounce is $V_{\rm b}=146.22\, V_{\rm Pl}$. As discussed earlier in \secref{sec:lqc}, 
here, $\Delta V$ refers to the expectation value of the dispersion in volume, $\langle\Delta V\rangle$.
In both of these cases, the initial state is constructed using \eqref{eq:wdwinteg}. 
In this paper, we have considered the evolution of symmetric initial states which have 
support on $V>0$, and have a very small amplitude near $V=0$. In principle one can 
consider states which have larger amplitude at or close to $V=0$. However, the part of 
the state close to $V=0$ contributes very little to the expectation values of the observables, 
while most of the contribution comes from the larger $V$ domain.
It is also noteworthy, that the relative spread in volume $\Delta V/V$ 
being greater than 1 does not imply that the support of the state extends 
to negative volume. It is also to be noted that the shape of the wavefunction is symmetric in 
the variable $\ln(v)$ which leads to high asymmetry around the peak when plotted against $v$ 
for widely spread states. Due to this, the part of the state that extends
to large volume contributes much more to the dispersion in volume $\Delta V$ than the 
part that extends to smaller volume. Thus, even though the dispersion are larger than 1
for the wide case, the state has support only on positive $V$. 
The bounce volume is computed after evaluating expectation values of the volume observable 
using quantum evolution equation in LQC. 
We consider semi-classical states, corresponding to these two cases, far 
from bounce and peaked on a classical trajectory on the expanding branch as the initial 
data. Then we numerically evolve the state backwards using the Chimera scheme. 
During this evolution, as the energy 
density grows and the spatial curvature becomes Planckian, the classical trajectory 
tends to fall into the big bang singularity. In the LQC evolution, the trajectory 
agrees with the classical one in the low curvature regime and follows it very closely 
as long as the curvature remains very small compared to Planckian value. 
During further backward evolution, as the energy density grows closer to the Planck 
density, the LQC trajectory starts to deviate from the classical trajectory, and unlike 
the classical theory, the LQC trajectory undergoes a bounce from a finite non-zero 
volume instead of collapsing into the singularity. This behavior of the LQC 
trajectories were obtained in Refs.~\cite{aps1,aps2,aps3} for semi-classical states 
initially sharply peaked on a classical trajectory, with large $p_\phi$. 

\Fref{fig:3dw1500} shows the evolution of a sharply peaked semiclassical 
state for $p_\phi=1500\, \sqrt{G}\hbar$ and $\Delta V/V=0.064$ where the amplitude of the 
wave-function is plotted with respect to the volume $V$ and emergent time 
$\phi$.  The initial state is constructed as described in~\secref{sec:lqc}. 
In the numerical 
simulations, the initial data is provided in the expanding branch and the 
evolution is carried backwards.  It is clear from the figure that the state 
remains sharply peaked throughout the evolution.  The bounce for the state 
occurs at $V_{\rm b}=1659.83\, V_{\rm Pl}$, when the energy density becomes 
$\rho \approx 0.408\,\rho_{\rm Pl}$. This value is slightly less than, yet 
very close to, the upper bound on the energy density, 
$\rcr=0.409\,\rho_{\rm Pl}$.

\begin{figure}[tbh!]
  \includegraphics[width=0.47\textwidth]{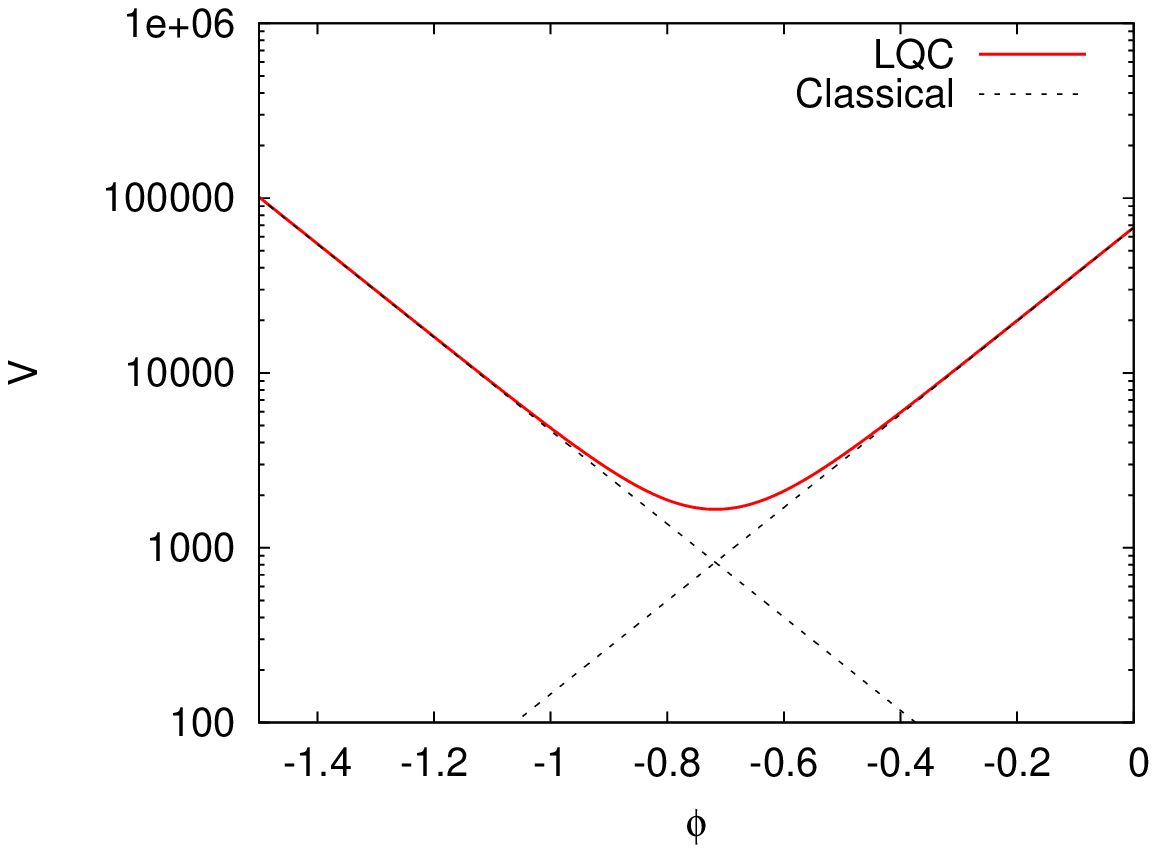}
  \includegraphics[width=0.47\textwidth]{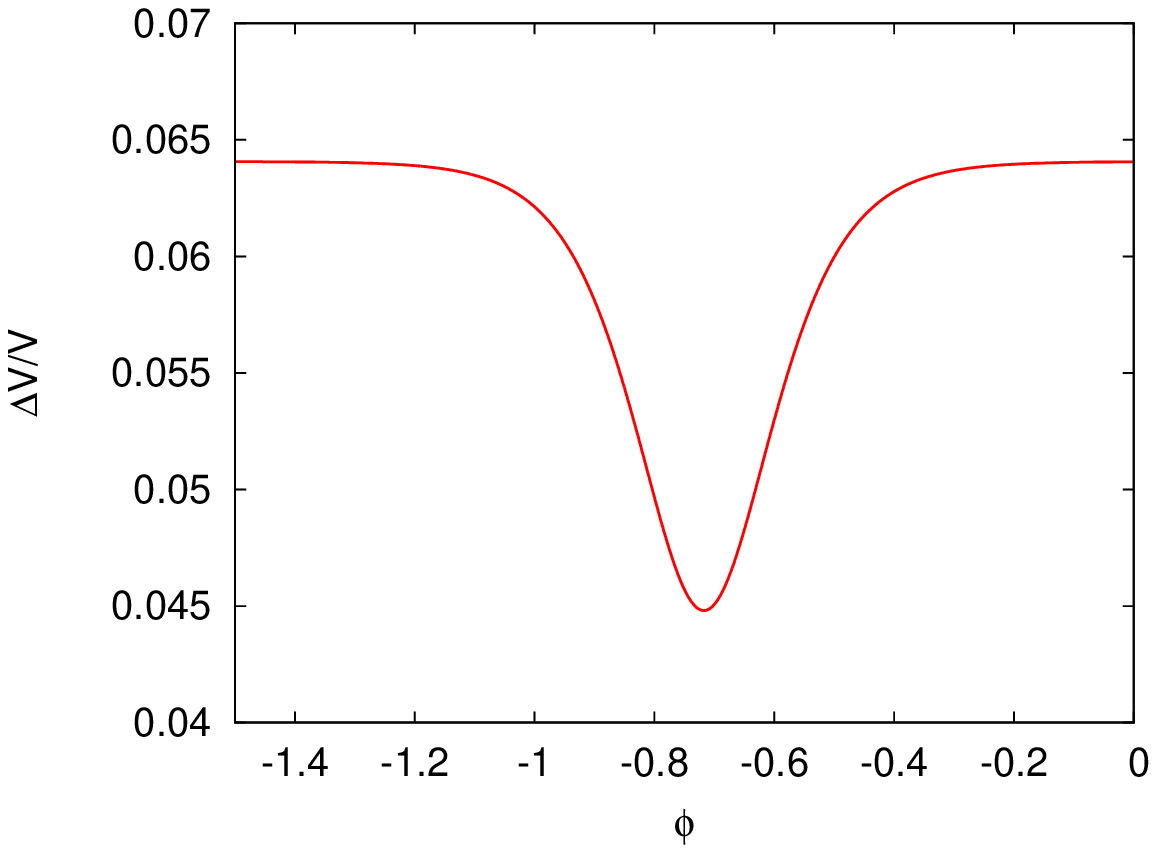}
  \caption{In this figure, the left plot shows the time evolution of the 
           expectation value of the volume and the right plot shows the 
           evolution of the relative volume dispersion of the state at each 
           time, for $p_\phi^*=1500\, \sqrt{G}\hbar$ and $\Delta V/V=0.064$. The 
           dashed curve in the left plot shows the classical trajectories.}
  \label{fig:1dw1500s121}
\end{figure}
\Fref{fig:1dw1500s121} shows the expectation value of the volume and the
relative dispersion in the volume as a function of $\phi$ 
for $p_\phi^*=1500\, \sqrt{G}\hbar$ and $\Delta V/V=0.064$.
The solid (red) curves show the LQC 
trajectory while the dashed (black) curves mark the two disjoint classical trajectories, one corresponding to the expanding branch and the other to the contracting branch for the same value of $p_\phi$.
It is 
straightforward to see from these plots that the LQC trajectory undergoes 
a non-singular bounce at finite volume. On the other hand, both of the 
classical trajectories continue to evolve towards the big bang.

As discussed in the previous sections, the numerical evolution of wide states with 
small $p_\phi$ is quite a challenge. Thanks to the Chimera scheme we can now 
perform such a numerical evolution very efficiently.  
\begin{figure}
  \includegraphics[width=\textwidth]{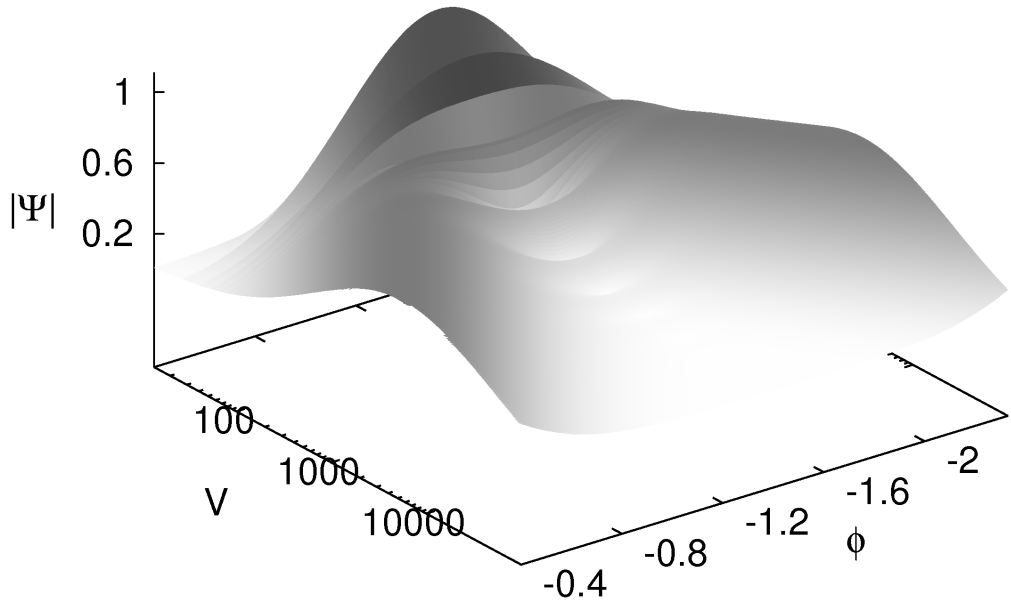}
  \caption{This figure shows the time evolution of the absolute value $|\Psi|$
           of a wide state corresponding to parameters
	   $p_\phi^*=20\,\sqrt{G} \hbar$, $\sigma=2.25$ and $\Delta V/V=6.44$. 
	   Far from bounce, both before and after, the state has
	   the form of a Gaussian, whereas it has clear non-Gaussian features 
	   in the vicinity of the bounce. Since the state is very spread out in $V$, we have 
	   used logscale on $V$-axis.} \label{fig:3dw20}
\end{figure}
\figref{fig:3dw20} shows the evolution of the absolute value of the state
$|\Psi|$ as function of $V$ and $\phi$
for the case of $p_{\phi}^*=20\,\sqrt{G} \hbar$, $\sigma=2.25$ and 
$\Delta V/V\approx6.44$.  Note that only a small range of $V$ is shown in the
figure. It is clear from the figure that far from the bounce the state has the form 
of a widely spread Gaussian whereas, in the vicinity of the bounce, the state 
has non-Gaussian features. The non-Gaussian feature of the state is more 
apparent in the first panel of \figref{fig:convergence4}, which shows a 
snapshot of the state at $\phi=-1.35$ close to the bounce. This 
behavior is very different from the evolution 
of the sharply peaked state (shown in \figref{fig:3dw1500}), where the 
state remains sharply peaked Gaussian throughout the evolution.
\begin{figure}[tbh!]
       \includegraphics[width=0.47\textwidth]{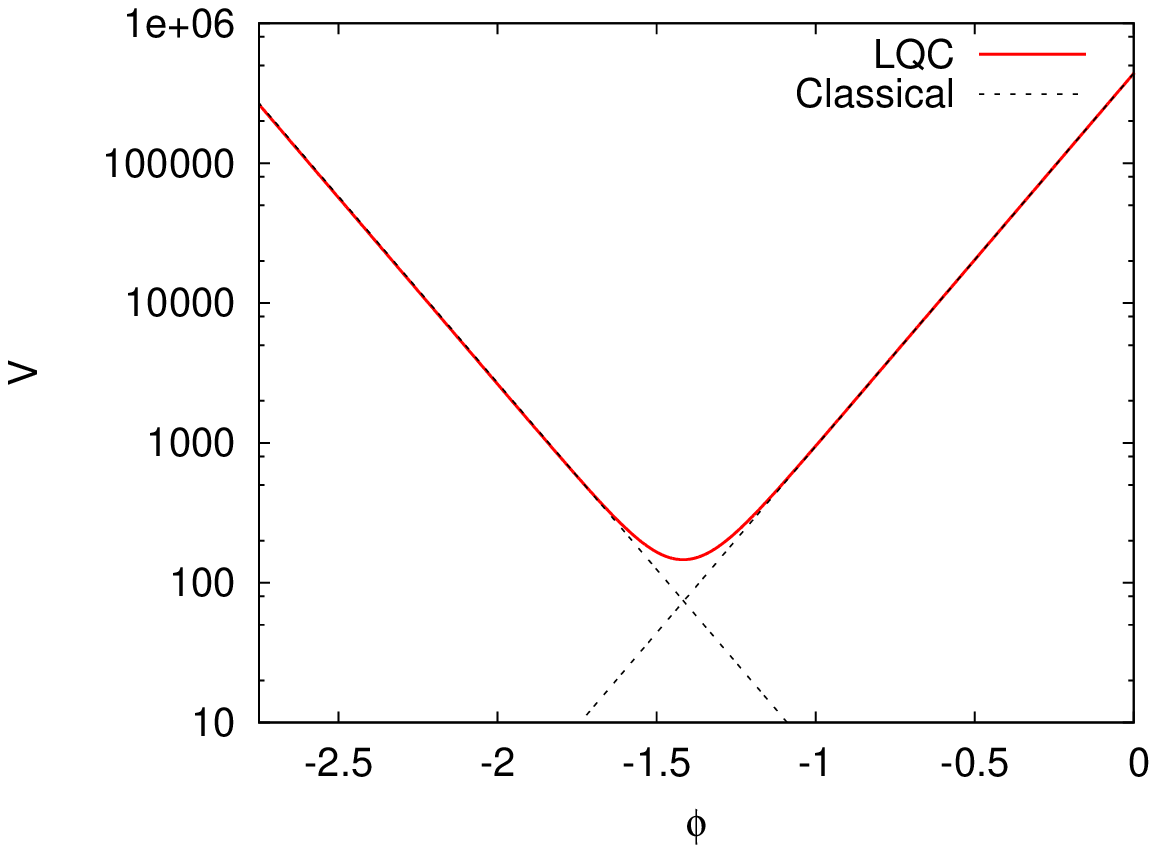}
       \includegraphics[width=0.47\textwidth]{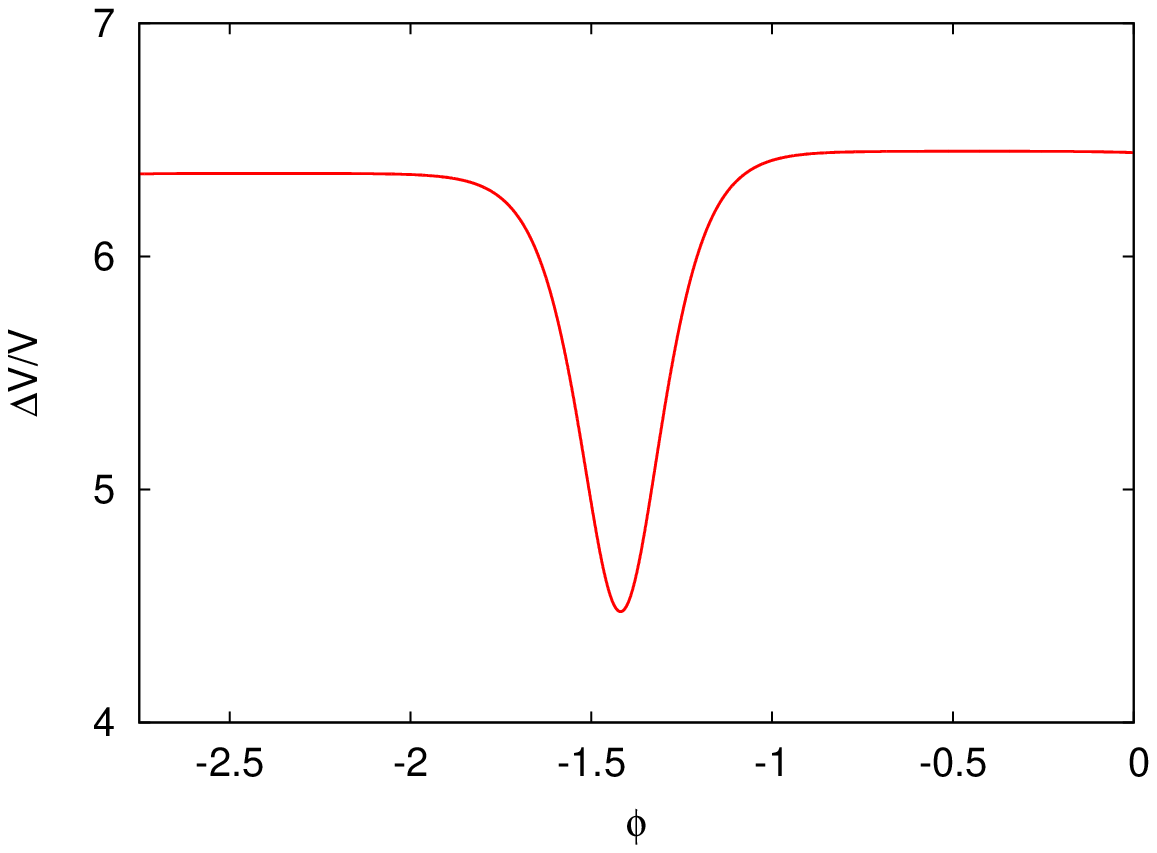}
\caption{The left plot shows the time evolution of the expectation value of the 
volume and the right plot shows the evolution of the relative volume dispersion
of the state as a function of $\phi$ for $p_\phi^*=20\,\sqrt{G} \hbar$, $\sigma=2.25$ and 
$\Delta V/V\approx6.44$. The dashed curve in the left 
plot shows the classical trajectories.}
\label{fig:1dw20s225}
\end{figure}
\figref{fig:1dw20s225} shows the expectation value of $V$ 
and $\Delta V/V$ with respect to $\phi$, corresponding to the evolution 
shown in \figref{fig:3dw20}. As can be seen from this figure, the 
qualitative features of the evolution trajectory of wide states are the 
same as those of the sharply peaked states.  
In the low curvature regime, the state follows the classical 
trajectory while in the Planck regime the corresponding classical trajectory 
reaches the big bang singularity while the LQC trajectory undergoes 
non-singular quantum bounce.  This happens at a much smaller (but still non
zero) volume compared to the sharply peaked state. There is also difference 
in the value of energy density at which the bounce occurs. In contrast to the 
previous case, this value turns out to be $\rho=0.009\,\rho_{\rm Pl}$, which 
is almost two orders of magnitude smaller than the upper bound on the 
energy density, $\rcr=0.409\,\rho_{\rm Pl}$.

In~\tref{tab:timings} we list representative run times for the 
$p_{\phi}^*=20\,\sqrt{G} \hbar$ and $\Delta V/V\approx6.44$ case using the Chimera scheme (for 
both the FD and DG implementations) on a 16 core 2.4 GHz 
Sandybridge workstation.  The first two columns give the grid size 
($n_{\mathrm{inner}}$) and run time of evolving only on the inner LQC grid, 
the next two columns give the outer grid size ($n_{\mathrm{outer}}$) and 
total run time for the finite difference Chimera scheme and the final two 
columns give the same information for the DG Chimera scheme. For both cases 
(FD and DG), $n_{\mathrm{outer}}$ was chosen so that the outer boundary was 
placed at $v_{\mathrm{outer}}=1.5\times 10^{19}$.

For the FD Chimera scheme, the resolution on the outer WDW grid is chosen to 
match the resolution of the LQC grid at the interface.  In the DG Chimera 
scheme we use 64th order elements and due to the much higher accuracy, we can 
use much lower resolution.  This is reflected in the much lower values for 
$n_{\mathrm{outer}}$ for the DG scheme in the table (compared to the FD 
scheme) and, though the computational cost per grid point is higher in the DG 
scheme than in the FD scheme, this the main reason for the corresponding much 
smaller run times.  We emphasize that the times listed for the FD and DG Chimera 
schemes include the time spent on the inner LQC grid.
\begin{table}
\caption{\label{tab:cost}Computational cost for various grid setups. The first
two columns lists the number of grid points ($n_{\mathrm{inner}}$) and the 
time spent on evolution on the inner LQC grid. The next 2 columns 
lists the number of grid points ($n_{\mathrm{outer}}$) on the outer WDW
grid and the total evolution time (on both the inner and outer grid) 
with the finite difference implementation of the Chimera scheme. The last 
2 columns lists the same quantities for evolution with the discontinuous 
Galerkin Chimera scheme. For both the FD and DG chimera schemes the grid is
set up so that the outer boundary is at $v_{\mathrm{outer}}=1.5\times 10^{19}$.
The timings are given in total walltime (seconds) when running on 16 cores on
a 2.4 GHz Sandybridge workstation.}
\begin{indented}
\item[]\begin{tabular}{@{}rrrrrr}
\br
\multicolumn{2}{c}{LQC grid} &\multicolumn{2}{c}{Chimera (FD)} & \multicolumn{2}{c}{Chimera (DG)} \\
$n_{\mathrm{inner}}$ & time (s) & $n_{\mathrm{outer}}$ &
  time (s) & $n_{\mathrm{outer}}$ &
  time (s) \\
\mr
  7,500 & 38.8 & 253,953 & 1117.6 & 8,125 & 143.5 \\
  15,000 & 87.4 & 497,508 & 9148.8 & 15,860 & 471.3 \\
  30,000 & 239.3 & 974,221 & 40339.0 & 31,005 & 1671.0 \\
  60,000 & 870.0 & --- & --- & 60,645 & 6160.0 \\
\br
\end{tabular}
\end{indented}\label{tab:timings}
\end{table}
The evolution time on the inner LQC grid reveals  $n_{\mathrm{inner}}^2$ scaling for large
$n$, and from this we can estimate that a pure LQC simulation with the outer
boundary at $v_{\mathrm{outer}}=1.5\times 10^{19}$ would take 
$3.4\times 10^{30}$ s $= 1\times 10^{23}$ years on this machine (if it had
enough memory).  For the Chimera scheme
the run time is reduced to about half a day for the FD implementation and 
less than about half an hour for the DG implementation (for the highest 
resolution listed in the table).  For the FD Chimera scheme we did not execute the computational run 
with $n_{\mathrm{inner}}=60,000$ since it would take a long time.  For the DG
Chimera scheme the run at $n_{\mathrm{inner}}=30,000$ gives accurate enough
results, and a higher resolution is not required for discussion of results. 
In summary, though most of the computational saving comes from the Chimera 
scheme, the computational cost was brought down by another order of 
magnitude through the use of advanced DG methods rather than a straightforward
FD implementation. 

We now comment on the cost estimates for an implementation of the Chimera 
scheme to anisotropic models. For simplicity let us consider the case of an 
LRS Bianchi-I model with a massless scalar field. In this spacetime, two of
the scale factors and their time derivatives are
equal, while the third scale factor is different.  Taking into account that the
computational cost of updating one grid point will increase (due to additional
evolved variables and derivative terms) and based on the current memory
and computational requirements for the 1+1 dimensional code, we can get rough
estimates of the requirements for a 2+1 dimensional code.  Here, the
number of grid points (i.e.\ memory requirements) would scale as $n^2$, 
as the spatial grid for LRS Bianchi-I spacetime would be two dimensional, and
the computational cost as $n^3$ (two powers of $n$ from the number of grid
points and one power of $n$ from the smaller step size), where $n$ is
the number of grid points in one dimension.  For the lower resolution case
in~\tref{tab:timings} with the outer boundary at 
$v_{\mathrm{outer}}=1.5\times 10^{19}$, we estimate that a 2+1 dimensional 
code would require of the order 50 Gb of memory and about 20,000 core hours.  
The next higher resolution would require about 200 Gb of memory and about 
160,000 core hours.  Assuming that we can write a well scaling code, such 
simulations could be done comfortably within a day (lowest resolution) or a 
week (higher resolution) running on 1,000 cores on a typical modern super 
computer.

\section{Tests of the Chimera scheme}\label{sec:testchimera}
In the previous section, we have discussed that the DG method is far
more efficient and computationally inexpensive compared to the FD method. For this reason, we will focus on the discussion of the DG method. We now describe the
concept of convergence for the Chimera scheme for the DG method, and  present detailed tests. There
are two major concerns regarding the accuracy of the numerical results. The first
is the effect of the interface between the LQC and WDW grid and the second is
due to the initial data being chosen as a WDW state with an additional phase (\eqref{eq:wdwinteg}). With regard to
the first issue, we have to choose
the location of $v_{\mathrm{interface}}$ large enough such that the WDW equations
are a good approximation to the LQC equations. We can control the errors
from the interface by performing runs with different values for 
$v_{\mathrm{interface}}$ and compare the results. As the resolution on
the WDW grid is chosen to match the resolution on the LQC grid in $v$-space,
increasing $v_{\mathrm{interface}}$ automatically results in higher resolution
(in $x$) on the WDW grid. Thus, increasing $v_{\mathrm{interface}}$ results
in higher accuracy due to two different factors. The WDW equations are better
approximations to the LQC equations, and the WDW equations are solved with
higher numerical resolution.  With regard to the second issue, we are
interested in setting up a simulation where the state starts to travel
inwards (towards smaller volume) as the evolution proceeds.  We will refer to
this as an ingoing mode.  Similarly
we refer to a mode that initially travels outwards (towards larger volume) 
as an outgoing mode.  Initial data constructed as
a pure ingoing WDW state will have a small (depending on where the state
is peaked and how wide it is) outgoing LQC mode. We can reduce the 
amplitude of this outgoing
mode by starting the simulation at a larger value of $\phi_{\mathrm{o}}$.
However, we demonstrate that, if the state is setup on the WDW grid, the interface 
between the LQC and WDW grids also helps by
acting as a filter that reflects the outgoing LQC mode.
\subsection{Convergence for a sharply peaked state}
\begin{figure}
  \includegraphics[width=0.98\textwidth]{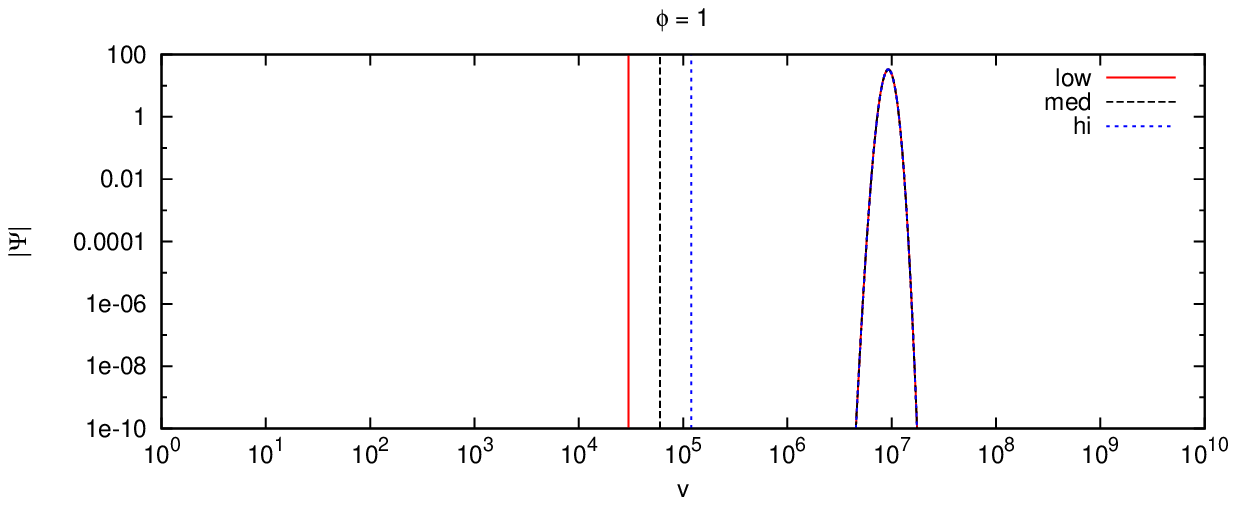}
  \includegraphics[width=0.98\textwidth]{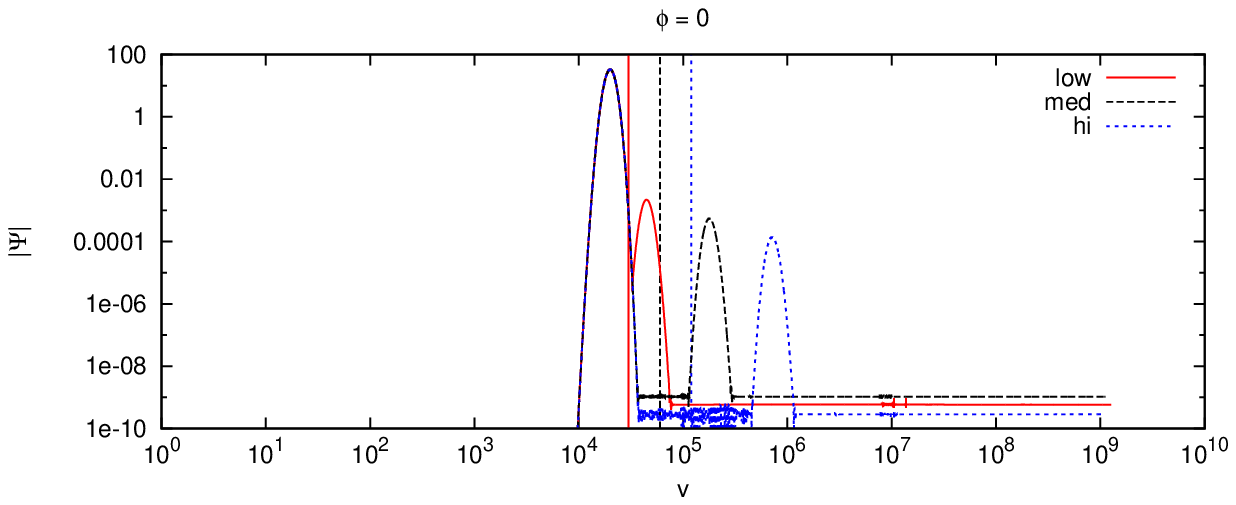}
  \caption{Log-log plots of the absolute value of the state as a function
    of the volume variable $v$.  Here snapshots of the evolution of
    a sharply peaked state with $p_{\phi}^*=1500\,\sqrt{G} \hbar$ and $\Delta V/V=0.064$
    at $\phi = 1$ (upper panel) and $\phi = 0$ (lower panel). 
    The solid (red), 
    dashed (black) and dotted (blue) curves are for the low 
    ($v_{\mathrm{interface}}=30,000$), medium 
    ($v_{\mathrm{interface}}=60,000$) and high
    ($v_{\mathrm{interface}}=120,000$) resolution runs, respectively.
    For clarity the thinner vertical lines shows the position of 
    $v_{\mathrm{interface}}$ for the three cases.}\label{fig:convergence1}. 
\end{figure}

\begin{figure}
  \includegraphics[width=0.98\textwidth]{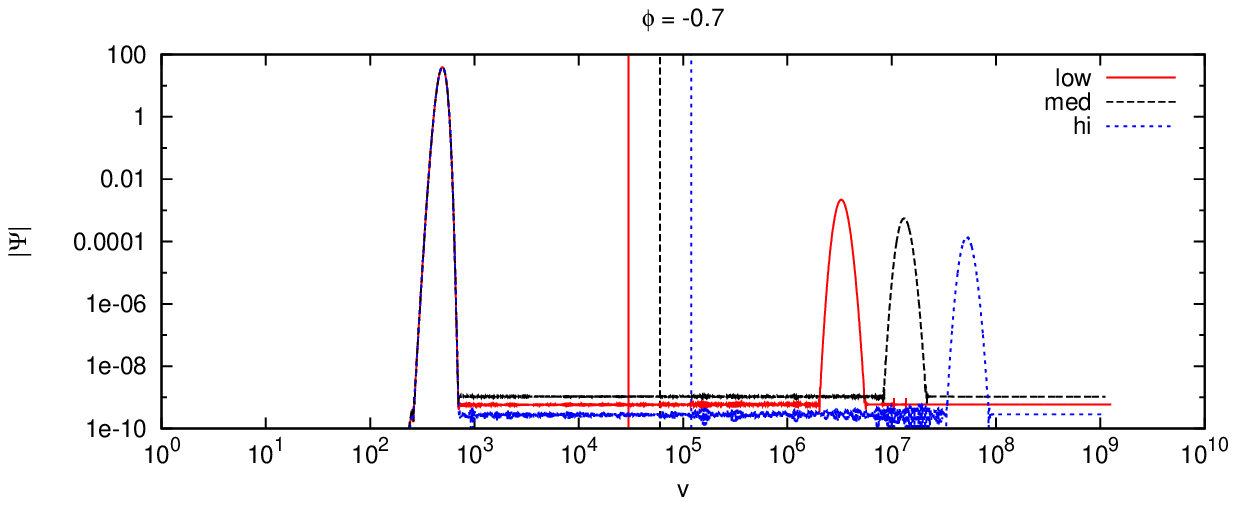}
  \includegraphics[width=0.98\textwidth]{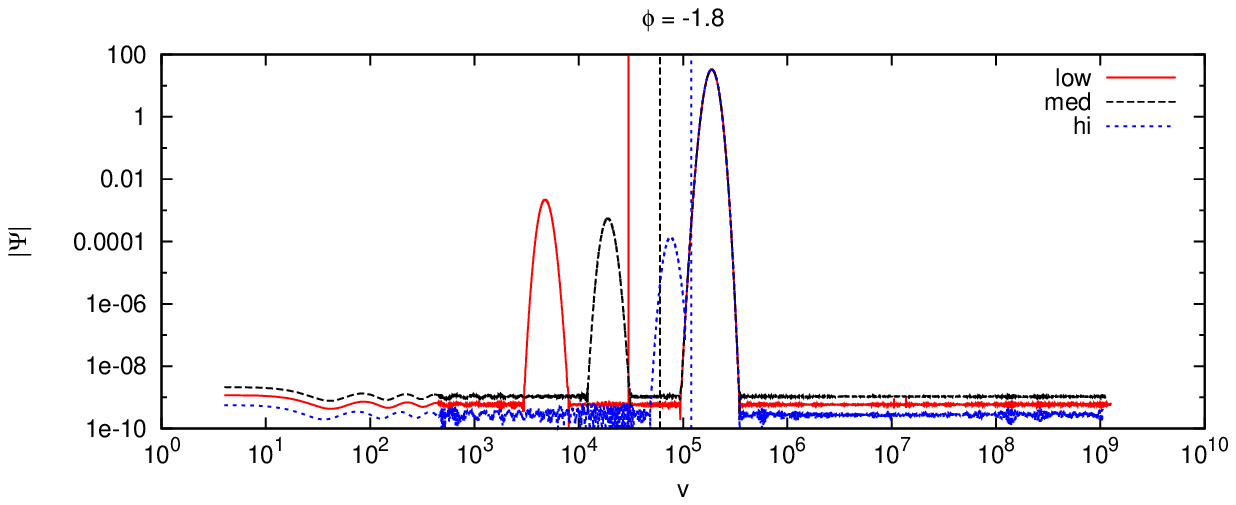}
  \caption{Log-log plots of the absolute value of the state as a function
    of the volume variable $v$.  Here snapshots of the evolution of 
    a sharply peaked state with $p_{\phi}^*=1500 \,\sqrt{G}\hbar$ and $\Delta V/V=0.064$
    at $\phi = -0.7$ (upper panel)
    and $\phi = -1.8$ (lower panel). The solid (red), 
    dashed (black) and dotted (blue) curves are for the low 
    ($v_{\mathrm{interface}}=30,000$), medium 
    ($v_{\mathrm{interface}}=60,000$) and high
    ($v_{\mathrm{interface}}=120,000$) resolution runs, respectively.
    For clarity the thinner vertical lines shows the position of 
    $v_{\mathrm{interface}}$ for the three cases.}\label{fig:convergence2}. 
\end{figure}

In figures~\ref{fig:convergence1} and~\ref{fig:convergence2} we show four 
snapshots of the evolution of a
sharply peaked state with $p_{\phi}^*=1500 \,\sqrt{G} \hbar$ and $\Delta V/V=0.064$ starting at
$\phi=1$.  The upper panel in~\figref{fig:convergence1} shows the initial state
at $\phi = 1$ while the lower panel shows the state at $\phi = 0$ when it has
just passed through the interfaces, denoted by vertical lines. We consider three 
different resolutions, corresponding to three different locations of the interface: 
the solid (red), dashed (black) and dotted (blue) curves correspond to the low, 
medium and high resolution respectively. It is clear from the figures that as the 
state passes through an interface, a part of the state is reflected from it. Properties of the 
reflected parts depend on the location of the interface. For example, in the higher 
resolution case, since the interface is further out, the reflection occurs earlier and 
the reflected part of the state has propagated to a larger volume than the low and 
medium resolution cases. Moreover, the amplitude of the reflected part of the 
state converges to zero to second order in $v_{\mathrm{interface}}$.  That is, 
when $v_{\mathrm{interface}}$ doubles, the amplitude decreases by a factor of 
four. The almost flat but noisy features trailing the state at around the $10^{-9}$
level (different at different resolutions) is caused by a combination of
roundoff and truncation errors\footnote{Finite precision in the 
representation of floating point numbers used in computers result in roundoff
error, while truncation errors are caused by using finite resolution in the 
approximation of derivatives.}. In~\figref{fig:convergence2}, the upper panel
shows the state at $\phi=-0.7$
close to the bounce which occurs at $\phi_{\mathrm{b}} = -0.71744$ at $V_{\mathrm{b}}=1659.83 V_{\mathrm{Pl}}$.  In this figure, the reflected part of the waves have propagated 
further to the right (towards larger $v$). The lower panel of~\figref{fig:convergence2} 
shows the state at $\phi=-1.8$ just after it has crossed the interface one more time.  
Again there is a reflected part of the state, now propagating to the left (towards 
smaller $v$).  The amplitude is again second order convergent in 
$v_{\mathrm{interface}}$ and since the state has to propagate to larger $v$
before hitting the interface in the high resolution case, the reflected
part of the state is delayed, as compared to the lower
resolution cases. This allows a larger time interval between the two 
reflections (first before bounce and the second afterwards) at the 
interface, in the case of high resolution. 
The first reflection is caused by the fact that the initial state
is set up as a purely ingoing WDW state.  It evolves cleanly until it hits the
interface.  As this purely ingoing WDW state is not a purely ingoing 
LQC state, it contains a small outgoing component.  This component can
not be propagated inwards and is reflected.  In this sense the
interface acts as a filter that ensures that the state on the LQC grid
is purely ingoing.  If the initial state was set up completely
contained within the LQC grid, we would see a small part of the state
immediately starting to move outwards.  Note that the outer boundary in this
simulation was placed at $v\approx 10^{9}$, allowing us to set up the state
at $\phi=1$ where it is peaked at $v\approx 10^7$.  This would have been
prohibitively expensive using only an LQC grid.
\subsection{Convergence for a wide state}
\begin{figure}
  \includegraphics[width=0.98\textwidth]{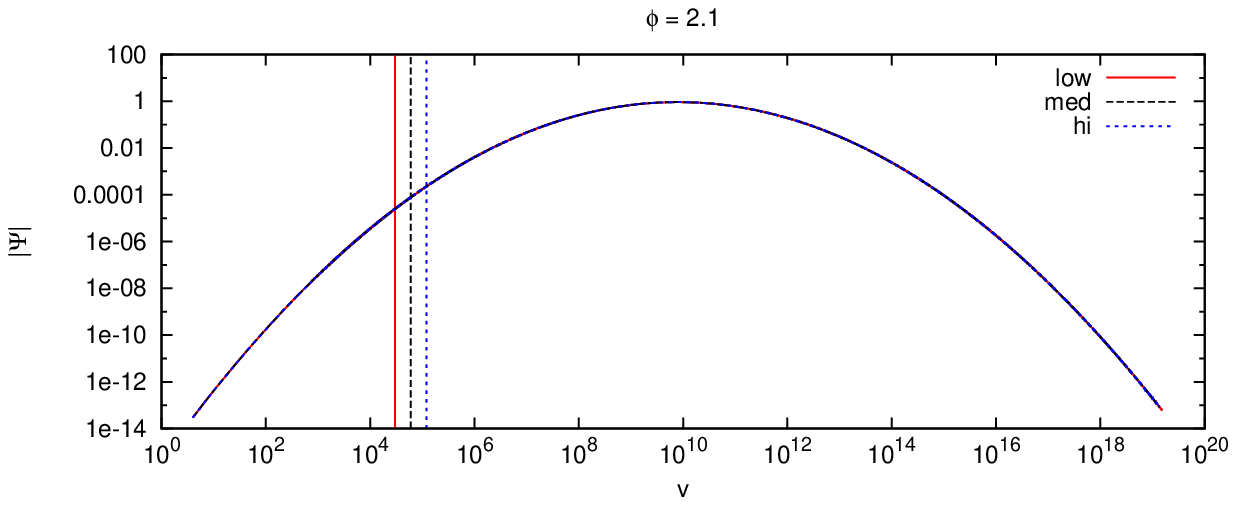}
  \includegraphics[width=0.98\textwidth]{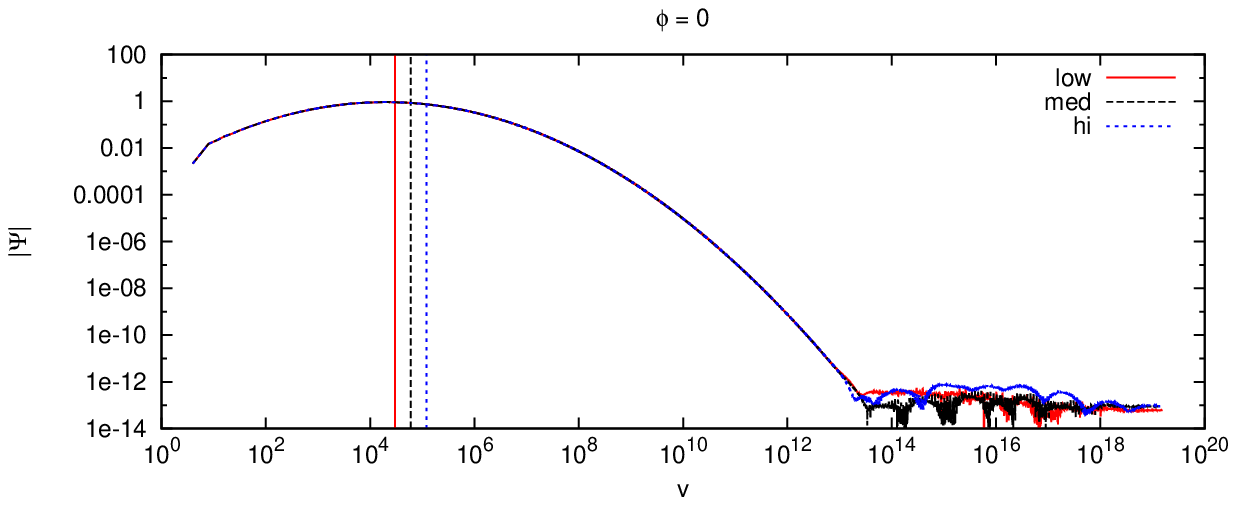}
  \caption{Log-log plots of the absolute value of the state as a function
    of the volume variable $v$.  Here snapshots of the evolution of a very wide 
    state with $p_{\phi}^*=20\, \sqrt{G}\hbar$ and $\Delta V/V\approx6.44$ at $\phi = 2.1$ 
    (upper panel) and $\phi = 0$ (lower panel). The solid (red), 
    dashed (black) and dotted (blue) curves are for the low 
    ($v_{\mathrm{interface}}=30,000$), medium 
    ($v_{\mathrm{interface}}=60,000$) and high
    ($v_{\mathrm{interface}}=120,000$) resolution runs, respectively.
    For clarity the thinner vertical lines shows the position of 
    $v_{\mathrm{interface}}$ for the three cases.}\label{fig:convergence3}. 
\end{figure}

\begin{figure}
  \includegraphics[width=0.98\textwidth]{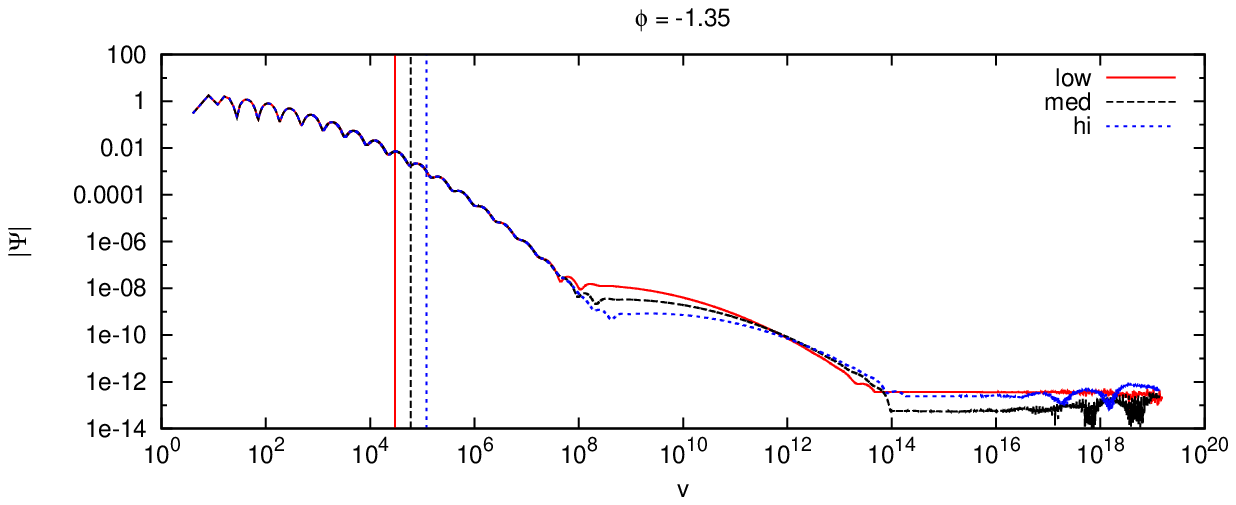}
  \includegraphics[width=0.98\textwidth]{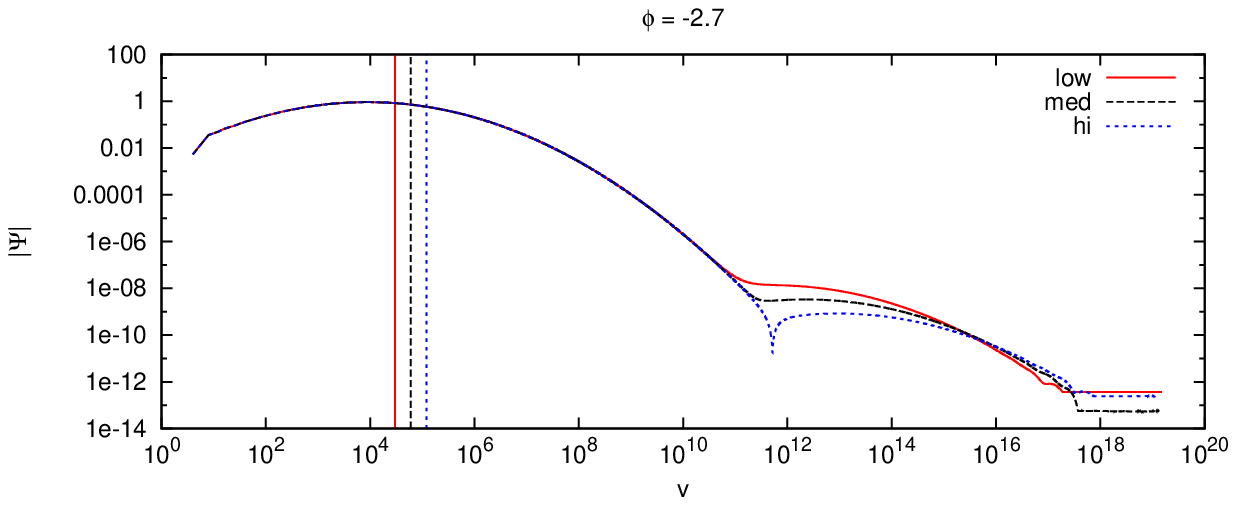}
  \caption{Log-log plots of the absolute value of the state as a function
    of the volume variable $v$.  Here snapshots of the evolution of a very wide 
    state with $p_{\phi}^*=20\,\sqrt{G} \hbar$ and $\Delta V/V\approx6.44$ at $\phi = -1.35$ 
    (upper panel) and $\phi = -2.7$ (lower panel). The solid (red), 
    dashed (black) and dotted (blue) curves are for the low 
    ($v_{\mathrm{interface}}=30,000$), medium 
    ($v_{\mathrm{interface}}=60,000$) and high
    ($v_{\mathrm{interface}}=120,000$) resolution runs, respectively.
    For clarity the thinner vertical lines shows the position of 
    $v_{\mathrm{interface}}$ for the three cases.}\label{fig:convergence4}. 
\end{figure}
In figures~\ref{fig:convergence3} and~\ref{fig:convergence4} we similarly show
four snapshots of the
evolution of a very wide state with $p_{\phi}^*=20 \, \sqrt{G}\hbar$ and $\Delta V/V\approx6.44$
starting at $\phi=2$.  The upper panel in~\figref{fig:convergence3} shows
the initial state at $\phi=2$.  As
in~\figref{fig:convergence1} and~\figref{fig:convergence2}, the vertical lines
indicate the position of the
interface boundary for each of the three different resolutions.  Note
that the state is so wide that we need to place the outer boundary at 
$v=10^{19}$ in order to contain the state within the numerical domain, and
that it has non-zero amplitude on the inner LQC domain.  This means that there
is a non-zero outgoing LQC mode from the beginning of the simulation.  However,
the amplitude is small and does not affect the measurement of the expectation
value of $V$ and its spread $\Delta V$.  The lower panel shows the state at 
$\phi=0$.
In the outer part of the grid the numerical solution is dominated by roundoff 
errors, while in the inner part of the grid part of the state has already
started to bounce.  The upper panel in~\figref{fig:convergence4} shows the
state at $\phi=-1.35$ close to the bounce which occurs at $\phi_{\mathrm{b}}=-1.4167$ at
$V_{\mathrm{b}}=146.219 V_{\mathrm{Pl}}$.  For $v>10^8$ it 
is  possible to see the part of the state that has been reflected off the 
interface boundary.  The amplitude of the reflection is much smaller compared 
to the case of the narrow state.  In this case the reflected amplitude is
about $10^{8}$ times smaller than the amplitude of the state.  As before,
the amplitude converges to zero at second order with the interface position
$v_{\mathrm{interface}}$. The lower panel shows the state at $\phi=-2.7$ well
after the peak has bounced.  At this point the shoulder of the reflection
propagates cleanly towards large $v$ ahead of the bouncing state.

It is clear from \figref{fig:convergence4} that, unlike 
the case of sharply peaked state whose shape always remains a
smooth and sharply peaked Gaussian in the evolution, the wide state with 
$p_{\phi}^*=20 \,\sqrt{G} \hbar$ and $\Delta V/V\approx6.44$ shows non-Gaussian 
features at $\phi\approx-1.35$ near the bounce. This behavior, as discussed in 
\sref{sec:results}, can also be seen in \fref{fig:3dw20} which shows a 3D evolution 
of the absolute value of the state, $|\Psi|$ with respect to $V$ and $\phi$.
However, as shown in the second panels 
of \figref{fig:convergence3} and \figref{fig:convergence4}, the shape of the wide
state far from bounce, both before (at $\phi=0$) and after the bounce (at $\phi=-2.7$), 
is a smooth Gaussian. 
Hence, far from the bounce 
the shape of the state is smooth 
for both the sharply peaked and widely spread states, while there are differences 
in the qualitative features of the shape of the wide states close to the bounce. 

\subsection{Convergence of $\Delta V/V$}
\begin{figure}
  \includegraphics[width=0.98\textwidth]{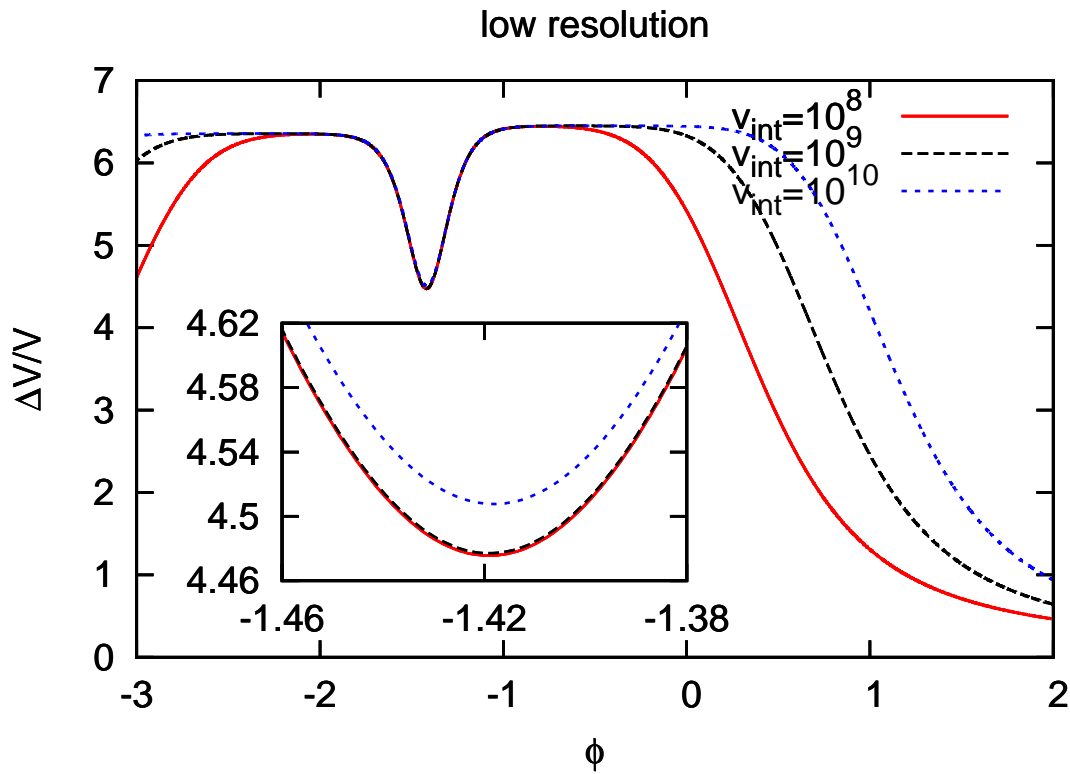}
  \caption{$\Delta V/V$ for 3 different integration radii:
  $v_{\mathrm{int}}=10^8$ (solid/red line), $v_{\mathrm{int}}=10^9$
  (dashed/green line) and $v_{\mathrm{int}}=10^{10}$ (dotted/blue line)
  for the case $p_{\phi}^*=20 \,\sqrt{G} \hbar$ and  $\Delta V/V\approx6.44$ at low resolution
  $v_{\mathrm{interface}}=30,000$. the inset shows in detail what 
  happens near the bounce.}\label{fig:deltavov_low}. 
\end{figure}

\begin{figure}
  \includegraphics[width=0.98\textwidth]{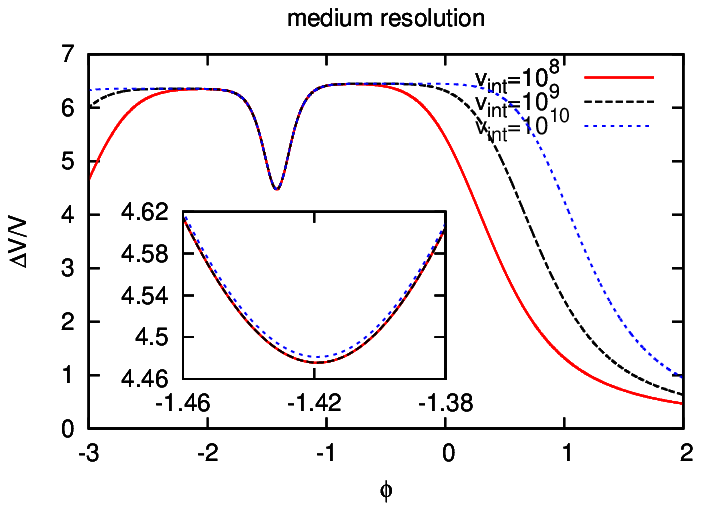}
  \caption{$\Delta V/V$ for 3 different integration domains:
  $v_{\mathrm{int}}=10^8$ (solid/red line), $v_{\mathrm{int}}=10^9$
  (dashed/green line) and $v_{\mathrm{int}}=10^{10}$ (dotted/blue line)
  for the case $p_{\phi}^*=20 \, \sqrt{G}\hbar$ and  $\Delta V/V\approx6.44$ at medium resolution 
  $v_{\mathrm{interface}}=60,000$. The inset shows in detail what 
  happens near the bounce.}\label{fig:deltavov_med} 
\end{figure}

\begin{figure}
  \includegraphics[width=0.98\textwidth]{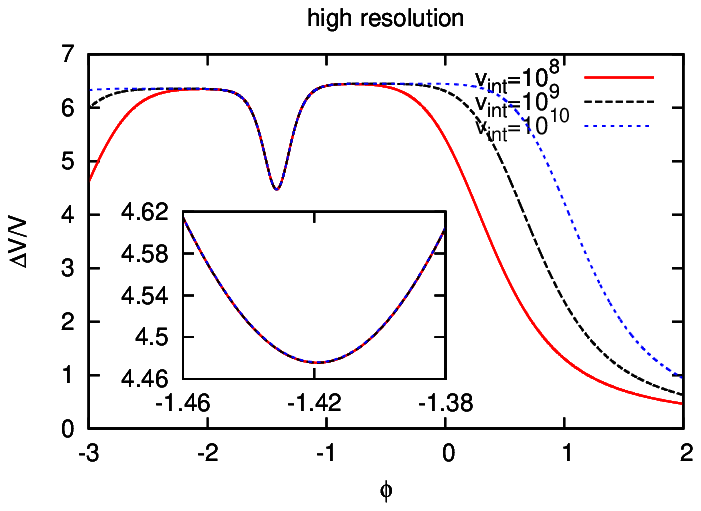}
  \caption{$\Delta V/V$ for 3 different integration domains:
  $v_{\mathrm{int}}=10^8$ (solid/red line), $v_{\mathrm{int}}=10^9$
  (dashed/green line) and $v_{\mathrm{int}}=10^{10}$ (dotted/blue line)
  for the case $p_{\phi}^*=20\, \sqrt{G}\hbar $ and  $\Delta V/V\approx6.44$ at high resolution 
  $v_{\mathrm{interface}}=120,000$. The inset shows in detail what 
  happens near the bounce.}\label{fig:deltavov_hi}. 
\end{figure}

When calculating the expectation value of the volume $V$ of the state and
its corresponding spread $\Delta V$, as given in~\eqref{eq:expectation} 
and~\eqref{eq:dispersion}, we need to integrate over the state.
As the integrand used to compute the expectation value of $v^2$ contains an extra 
factor of $v$, it is clear from~\figref{fig:convergence3} that we would get 
contributions from the roundoff error (noise at the $10^{-13}$ level contributes
significantly when multiplied by $10^{19}$) if the integration is extended
all the way to the outer boundary.  For this reason we need to choose a
volume $v_{\mathrm{int}}$ at which the integration is stopped.  On one hand, if 
we choose $v_{\mathrm{int}}$ to be too small, we may not have the whole state 
contained in the integration domain.  On the other hand, if we choose it too large 
we may get spurious contributions from the roundoff error noise or from the reflected 
part of the state.  In fact, in many cases, it is impossible to choose a value for 
$v_{\mathrm{int}}$ that is satisfactory for the whole simulation.  We illustrate this 
in figures~\ref{fig:deltavov_low}, \ref{fig:deltavov_med}
and~\ref{fig:deltavov_hi} for the extreme case of $p_{\phi}^*=20 \, \sqrt{G}\hbar$ and $\Delta V/V\approx6.44$ 
at low, medium and high resolutions respectively.
In all cases, the plots show the calculated value of $\Delta V/V$ as function
of $\phi$ using three different values of $v_{\mathrm{int}}$.  The 
solid (red) curve is for $v_{\mathrm{int}}=10^8$, the dashed (green) 
curve is for $v_{\mathrm{int}}=10^9$ and the dotted (blue) curve is 
for $v_{\mathrm{int}}=10^{10}$.  At late times, when the state is
peaked at large volume (the WDW regime), $\Delta V/V$ should be constant.
However, in this regime the state is not completely contained within the
integration volume and we get different values for $\Delta V/V$ for the
three different values of $v_{\mathrm{int}}$.  As the state
contracts further (as $\phi$ decreases) it becomes small enough that first 
$v_{\mathrm{int}}=10^{10}$, then $v_{\mathrm{int}}=10^9$ and
finally $v_{\mathrm{int}}=10^8$ is large enough to contain the state.
After this we get good agreement between the calculations of $\Delta V/V$
using the three different values of $v_{\mathrm{int}}$ through
the bounce (at $\phi=-1.42$ in this case) and for some time afterwards until
the state expands and become large enough that it no longer is contained within
the integration volume.  A close up of the near bounce region at low resolution
(see the inset in~\figref{fig:deltavov_low}), however, reveals some small
differences.  A small difference is visible between the 
$v_{\mathrm{int}}=10^8$ and $v_{\mathrm{int}}=10^9$ curves
while the $v_{\mathrm{int}}=10^{10}$ curve shows a clear deviation from
the other curves.  These are caused by the reflected part of the state of the 
interface between the inner LQC grid and the outer WDW grid seen
in~\figref{fig:convergence3}.  However, as the interface between the two grids
are moved out (and the resolution is increased), the amplitude of the 
reflection decreases.  Thus at the medium resolution (see the inset 
in~\figref{fig:deltavov_med}), the $v_{\mathrm{int}}=10^8$ and 
$v_{\mathrm{int}}=10^9$ curves agree very nicely, while the 
deviation of the $v_{\mathrm{int}}=10^{10}$ curve has decreased 
significantly compared to the low resolution case.  Finally in the
high resolution case (see the inset in~\figref{fig:deltavov_hi}) all three
curves agree nicely all the way through the bounce.  Thus we conclude that
the interface between the grids is far enough out in the ``classical'' region
that any artifacts introduced by the Chimera scheme do not affect the
results.  For different initial data parameters, the optimal choice of 
$v_{\mathrm{int}}$ will be different, so in each case we have to
perform simulations at different resolutions and examine the plots of 
$\Delta V/V$ for different $v_{\mathrm{int}}$.  For the sharply peaked
state shown in \figref{fig:convergence1} and \figref{fig:convergence2}, for
example, we find that 
$v_{\mathrm{int}}=10^5$ is sufficient. Even though the roundoff noise
is in the integration volume at the bounce, it is at a sufficiently low level
that it does not adversely affect the value of $\Delta V/V$.  Also the
reflected wave in that case (of much larger amplitude than the really wide
state) leaves the integration volume as early as $\phi\approx 0$ in the high
resolution case.

As the integrand for the expectation value of $v$ in~\eqref{eq:expectation}
only contains one factor of $v$, the errors for the expectation value are much
smaller than for the dispersion.  Thus convergence of $\Delta V/V$ implies
convergence of the expectation value of $V$ as well.

\subsection{Comparison with pure LQC evolution}
As a final test, we want to compare the Chimera scheme with a pure LQC evolution, i.e.\ with
the outer WDW grid turned off. As the wide state requires the outer boundaries
to be placed at $v=1.5\times 10^{19}$, it is unfeasible to make a comparison in this case. So for the comparison we use the sharply 
peaked state shown in figures~\ref{fig:convergence1} and~\ref{fig:convergence2}.
Those simulations where started at $\phi_o=1$ where the state is peaked at
$v\approx 10^7$, requiring the outer boundary to be located at at least
$v_{\mathrm{outer}}=2\times 10^7$. Using the results of runs in Table 1, the estimated time to complete such a simulation turns out to be approximately 70
days on a workstation. Thus, this simulation is too expensive to do with a pure LQC grid.  Therefore, to do the comparison we take the following strategy. We perform three pure LQC simulations with 
different starting times $\phi_o = (0, 0.1, 0.2)$ with 
$v_{\mathrm{outer}}=240,000$ and compare with the Chimera runs.  In these three
cases the state is initially peaked at $v=20,000$, $37,000$ and $70,000$,
respectively.  In all these cases, states are so sufficiently far inside the outer boundary, 
that we avoid any outer boundary problems.

\begin{figure}
  \includegraphics[width=0.98\textwidth]{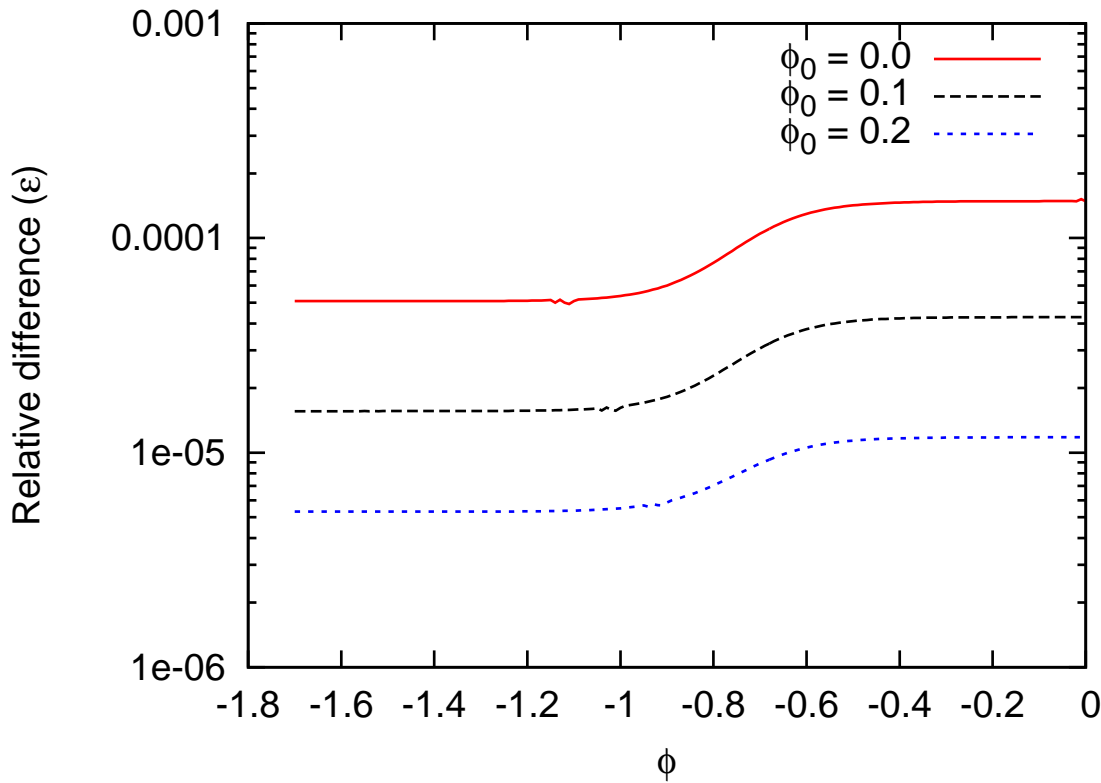}
  \caption{The relative difference 
    $\varepsilon=|\langle\hat{v}\rangle_{\mathrm{LQC}}-
               \langle\hat{v}\rangle_{\mathrm{Chimera}}|/
               \langle\hat{v}\rangle_{\mathrm{Chimera}}$ for different 
    starting $\phi_0$ as function of $\phi$ for the sharply peaked state 
    $p_{\phi}^*=1500 \,\sqrt{G} \hbar$ and  $\Delta V/V=0.064$. The solid (red) curve is for 
    $\phi_0=0$, the dashed (black) curve is for $\phi_0=0.1$ and the dotted
    (blue) curve is for $\phi_0=0.2$}\label{fig:nonhybrid}.  
\end{figure}
In~\figref{fig:nonhybrid} we plot the relative difference in the expectation
value of $v$ between the pure LQC simulations and the Chimera simulation as
function of $\phi$ for the three different starting times.  For $\phi_o=0$ the
relative difference is of order $10^{-4}$.  This difference is due to the
fact that the initial state is setup as a solution to the WDW equations and
therefore is not a pure ingoing mode for the LQC equations. We use the term
ingoing in the same sense here as we did in the beginning of \secref{sec:testchimera}.
As $\phi_o$
increases, the state is initially peaked at larger $v$ and the WDW initial
state is a better approximation to the pure ingoing LQC state.  Therefore
we see the relative difference decreases for larger values of $\phi_o$, and
for $\phi_o=0.2$ the difference is of order $10^{-5}$.  In fact the decrease
in relative difference shown in the figure is consistent with second order 
convergence in the starting location (in $v$) of the peak of the state and
is also consistent with the second order convergence in the amplitude of the 
reflected part of the state off the interface boundary when the Chimera scheme
is used.  This again shows the advantage of being able to use a large domain,
allowing us to setup the initial data peaked at a large enough volume that 
the errors made in constructing the state as a WDW state is negligible.

\section{Discussion}\label{sec:discussion}

In this article we have presented an efficient numerical technique to study 
the evolution of states in LQC. Previously, numerical simulations in isotropic 
LQC were performed with a massless scalar field as the matter source with
a large value of the field momentum for sharply peaked states. All such 
simulations show the existence of a quantum bounce at an energy density which
is very close to the maximum value of the energy density, 
$\rho_{\rm max} \approx 0.41 \rho_{\mathrm{Planck}}$, computed from an 
exactly solvable model  (with lapse $N = V$) \cite{acs}. Though the robustness 
of the bounce has been verified in various models, it was so far not observed 
for states with small field momentum and with very wide spreads, or for 
non-Gaussian states. Such states can be thought of as corresponding to  
universes which are more `quantum' compared to the states previously 
considered in numerical simulations in LQC. Using the results of \cite{acs}, 
it has recently been suggested that a certain class of squeezed states may 
lead to a quantum bounce \cite{montoya_corichi2} at a much lower energy 
density. Further, a detailed study of states which have wide spread, small field 
momentum and non-Guassian features is critical in order to understand the 
validity of the effective spacetime description in LQC \cite{psvt}.
However, testing the robustness of  new physics in LQC for such kinds of 
states using numerical simulation is a computationally challenging task, 
which is extremely difficult to perform with the existing techniques. These 
difficulties are further aggravated in the case of anisotropic models due to
the increased number of spatial dimensions (leading to much larger 
computational domains). In addition, certain non-trivial properties of the 
anisotropic quantum difference equation, which are absent in the isotropic
model, complicates the implementation of numerical codes.

In order to tackle these computational challenges, we have proposed and 
implemented a hybrid numerical technique, Chimera, which uses a combination
of LQC and WDW computational grids. This scheme is based on a crucial property 
of the quantum Hamiltonian constraint in LQC -- that it is approximated
extremely well by the WDW Hamiltonian constraint at large volumes.
The inner grid, corresponding to a lower volume range, is a uniform discrete
grid based on the discreteness given by LQC. It is taken to be reasonably
large such that all the non-trivial features of the physics of LQC is contained
within this domain, and there is little difference between the LQC and WDW 
evolution at the outer boundary of the inner grid. The outer grid, on the other hand, is a
grid with a continuum limit on which we solve the WDW equation. Here we have
utilized the fact that the isotropic LQC difference equation can be approximated by a
second order finite differencing approximation to the WDW equation in the
large volume limit. Using advanced techniques for solving PDE's, we can
explore efficient ways to evolve the state on the outer grid. We studied two
such methods: a finite difference implementation, and a discontinuous Galerkin
method. 
We find that discontinuous Galerkin method turns out to be extremely efficient.
Evolutions with widely spread states which would have taken $10^{23}$ years can
now be performed in a few hours on a modern workstation. We further performed
various convergence tests with the discontinuous Galerkin method, verified
crucial properties of the evolutions of sharply and widely peaked states and
studied the behavior of the relative fluctuations in comparison with pure LQC
evolutions.
Our analysis shows that the Chimera scheme proves to be very advantageous both in
terms of the accuracy of the results and the computational cost. More
importantly, with the scheme presented here, we are able to perform simulations
with a very wide variety of states. Using the run time estimates for isotropic
models, the Chimera scheme also promises a very significant reduction in
computational costs for the anisotropic models. The computational costs in
these models scales as $n^3$ for LRS Bianchi-I spacetimes, and as $n^4$ for
Bianchi-I models with a massless scalar field, making them practically very
difficult to study even with supercomputers. Our estimates show that using
the Chimera scheme, many of these simulations can be performed in less than a
week on current super computers. Thus, the Chimera scheme promises to be an 
important tool to explore so far unknown regimes of LQC in various models.

The current article is the first in a series of papers focused towards building
an efficient and portable numerical infrastructure for the simulations of
various cosmological models in the paradigm of LQC. Here, we have only
provided the details of the numerical properties of the scheme using sharply
peaked and widely spread Gaussian states for the spatially flat isotropic
model (with lapse $N=1$). Our results can then be viewed as a generalization
of earlier results in \cite{aps3} where simulations were carried out for
sharply peaked states for large $p_\phi$. In an upcoming paper, we will use
the Chimera scheme to understand the validity of the effective spacetime 
description
in detail for widely spread states by exploring regions much closer to Planck
volume than considered so far in the literature \cite{dgs2}.
As discussed in the paper, our scheme is not limited to simulations of
Gaussian states in isotropic loop quantum cosmology. We have used the Chimera
scheme to perform numerical evolutions of various other non-Gaussian states,
such as squeezed states and other states with nontrivial features \cite{dgms}.
Going beyond the massless scalar models, the Chimera scheme has also been
used to perform numerical simulations in LQC with negative potentials giving
rise to cyclic models of the universe \cite{dgms2}. In ongoing work, we
are implementing the Chimera method to study the numerical evolution of 
models with more degrees of freedom using high performance computing resources.

\ack
The authors would like to thank Steve Brandt, Frank L\"offler,  Miguel Megevand and Jorge Pullin
 for helpful comments and many fruitful discussions. This work is supported by a grant from John Templeton Foundation and by NSF grant PHYS1068743. The opinions expressed in this publication are those of authors and do not necessarily reflect the views of John Templeton Foundation. BG is also partially supported by the Coates Scholar Research Award of Louisiana State University.

\appendix

\section{Discontinuous Galerkin methods}
\label{ap:dg}
\begin{figure}
\begin{center}
\includegraphics[width=0.7\textwidth]{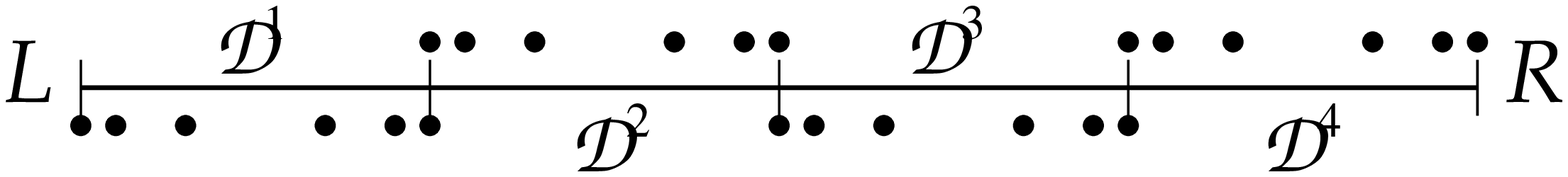}
\end{center}
\caption{This figure illustrates the way the numerical domain $\Omega=[L,R]$ is
         split into 4 non-overlapping 5th order elements
         ${\cal D}^1, \ldots, {\cal D}^4$. The black dots above (for
         ${\cal D}^1$ and ${\cal D}^3$) and below (for ${\cal D}^2$ and
         ${\cal D}^4$) the coordinate line shows the location of the 6 nodes
         inside each 5th order element. Notice that the nodes are not
         distributed evenly inside the element and that the boundary nodes on
         neighboring elements have the same coordinates.}\label{fig:dgmesh}
\end{figure}

In this appendix we describe the main ideas behind Discontinuous Galerkin (DG)
methods. The notation and discussion is borrowed heavily
from~\cite{Hesthaven:2007:NDG:1557392} which we recommend for further reading.
To understand DG methods, it is easiest to look at the simplest possible PDE, i.e.\ the
1-d scalar advection equation
\begin{equation}
  \frac{\partial u}{\partial t}+\frac{\partial f(u)}{\partial x} = 0,
  \mbox{\hspace*{2em}} x \in [L,R] = \Omega, \label{eq:continuum}
\end{equation}
where $f(u) = a u$ and $a$ is a constant. The numerical
domain $\Omega$ is approximated by $K$ non-overlapping elements
${\cal D}^k$
\begin{equation}
  \Omega\simeq\Omega_h=\bigcup_k^K{\cal D}^k,
\end{equation}
where $x\in [x_l^k,x_r^k]={\cal D}^k$, and where $x_r^k=x_l^{k+1}$
(see~\figref{fig:dgmesh}).
Within each element, the local solution is approximated as a polynomial of order 
$N=N_p-1$:
\begin{equation}
x\in {\cal D}^k: u_h^k(x,t) = \sum_{n=1}^{N_p} \hat{u}_n^k(t)\psi_n(x) =
\sum_{i=1}^{N_p}u_h^k(x_i^k,t)\ell_i^k(x). \label{eq:expansion}
\end{equation}
Here $u_h^k(x,t)$ denote the numerical approximation to the solution $u(x,t)$
within element $k$. In the first sum the numerical approximation is expressed
in terms of an expansion in a polynomial basis, where $\psi_n(x)$ are the basis
functions and $\hat{u}_n^k$ are the coefficients. This is called the modal
form. In the second sum the numerical approximation is expressed in terms of
a sum of interpolating polynomials. Here a set of $N_p$ distinct nodes
$x_i^k$ (see~\figref{fig:dgmesh}) are defined within each element and the
interpolating polynomials $\ell_i^k(x)$ satisfies
$\ell_i^k(x_j^k)=\delta_{ij}$. This is
called the nodal form. In the following, we will only consider the nodal form.
The global solution is then approximated by the piecewise $N$-th order
polynomial
\begin{equation}
u(x,t)\simeq u_h(x,t) = \bigoplus_{k=1}^K u_h^k(x,t),
\end{equation}
i.e.\ the direct sum of the $K$ local polynomial solutions $u_h^k(x,t)$.
Note that the global solution is multi-valued at element boundaries.

We finally define the numerical residual
\begin{equation}
{\cal R}_h(x,t) = \frac{\partial u_h}{\partial t}+
                  \frac{\partial f(u_h)}{\partial x}
\end{equation}
by inserting the numerical approximation to the solution $u_h(x,t)$ into the
continuum~\eqref{eq:continuum}.  Since the numerical solution has a
finite number of degree of freedom, the residual ${\cal R}_h(x,t)$ can not be
zero everywhere (unless the continuum solution itself is an $N$th order
polynomial).  To quantify how small the residual is, it is natural to use
a suitable defined inner product and norm.
Within element $k$, we define the local inner product and norm as
\begin{equation}
\lprod{u}{v}{k} = \int_{{\cal D}^k} uv\, dx,
                  \mbox{\hspace*{1em}} \|u\|_{{\cal D}^k}^k = \lprod{u}{u}{k},
\end{equation}
and with this we define the global inner product and norm as
\begin{equation}
\gprod{u}{v} = \sum_k^K\lprod{u}{v}{k}, 
               \mbox{\hspace*{1em}} \|u\|_{\Omega, h} = \gprod{u}{u}.
\end{equation}
In a Galerkin method the choice is then made to require that the residual is
orthogonal to all the basis functions in which the numerical solution is
expanded, i.e.\
\begin{equation}
\lprod{{\cal R}_h}{\psi_n}{k}=\int_{{\cal D}^k}{\cal R}_h(x,t)\psi_n(x)\, dx =
                              0, \mbox{\hspace*{1em}} 1\le n\le N_p.
\end{equation}
This is a local statement on each element and does not connect the solution
between neighboring elements. Since the interpolating polynomials can be
expanded uniquely in the basis functions an equivalent set of conditions can
be written in nodal form
\begin{equation}
\lprod{{\cal R}_h}{\ell_i^k}{k}=\int_{{\cal D}^k}{\cal R}_h(x,t)\ell_i^k(x)\, dx =
                              0, \mbox{\hspace*{1em}} 1\le i\le N_p.
  \label{eq:residualnorm}
\end{equation}
\Eref{eq:residualnorm} results in $N_p$ equations for the $N_p$
unknowns $u_h^k(x_i^k,t)$ within each element.
Using integration by parts we find
\begin{equation}
\lprod{{\cal R}_h}{\ell_i^k}{k}=\int_{{\cal D}^k}\left (\frac{\partial u_h^k}
{\partial t}-f(u_h^k)\frac{\partial \ell_i^k}{\partial x}\right )\, dx+ 
\left [ f(u_h^k) \ell_i^k\right ]_{x_l^k}^{x_r^k}, \label{eq:intbypart}
\end{equation}
where the surface term can be used to glue the solution in neighboring elements
together.  This can be achieved by replacing $f(u_h^k)$ in the boundary term
in~\eqref{eq:intbypart} by a numerical flux $f^*$ that depends
on the solution on both sides of the boundary.  This leads to the so called
weak form of the equations
\begin{equation}
\int_{{\cal D}^k}\left (\frac{\partial u_h^k}{\partial t}-
   f(u_h^k)\frac{\partial \ell_i^k}{\partial x}\right )\, dx =
-\left [ f^* \ell_i^k\right ]_{x_l^k}^{x_r^k},
\mbox{\hspace*{1em}} 1\le i\le N_p,
\end{equation}
where $f^*$ at the right boundary of element $k$ is a suitable function of both
$u_h^k(x_r^k)$ (from element $k$ and $u_h^{k+1}(x_l^{k+1})$ (from element
$k+1$).
Using integration by parts on the weak form of the equations we obtain the
strong form of the equations
\begin{equation}
\int_{{\cal D}^k} {\cal R}_h(x,t)\ell_i^k(x)\, dx  = 
   \left [(f(u_h^k)-f^*)\ell_i^k\right ]_{x_l^k}^{x_r^k},
   \mbox{\hspace*{1em}} 1\le i\le N_p. \label{eq:strong}
\end{equation}
It turns out that the choice of $f^*$ is not unique.  Often it can be tailored to be optimal
for the set of equations being considered.  An often used expression for the
 numerical flux that works well in many cases is the Lax-Friedrichs flux
along the normal $\hat{n}$
\begin{equation}
f^*=f^{\mathrm{LF}}(a,b) = \frac{f(a)+f(b)}{2}+\frac{C}{2}\hat{n}(a-b),
\end{equation}
where $(a,b)$ represents internal and external values, respectively and
\begin{equation}
0\le C\le\max \left |\frac{\partial f}{\partial u}\right |
\end{equation}
where the maximum has to be taken over all nodes in the element.  For the
simple advection equation it can be proven that this choice of flux leads
to a stable scheme.  For $C=0$
we have a symmetric flux that just depends on the average flux on either
side of the element boundary (hence the name).  For positive values of $C$ the
 jump in the solution across the element boundary is taken into account.
For $f(u)=a u$ and $C=|a|$ the Lax-Friedrich flux becomes an upwinding flux
in the sense that information is used only in the direction in which it is
traveling.
Inserting now the expansion of the numerical approximation in interpolating
polynomials from~\eqref{eq:expansion} into the strong form of the
equations~\eqref{eq:strong} and rewriting in matrix form we find
\begin{equation}
{\cal M}^k\frac{d}{dt}\bm{u}_h^k +
{\cal S}^k a \bm{u}_h^k = 
\left [(a u_h^k-(a u_h)^*)\bm{\ell}^k (x)\right ]_{x_r^k}^{x_l^k},
\end{equation}
where we have defined the solution vector $\bm{u}_h^k$ and interpolating
polynomial vector as
\begin{equation}
\bm{u}_h^k = \left [ u_h^k(x_1,t), \ldots, u_h^k(x_{N_p},t)\right ],
\mbox{\hspace*{2em}}\bm{\ell}^k (x) = \left [ \ell_1^k(x), \ldots, 
\ell_{N_p}^k(x)\right ]
\end{equation}
and the mass matrix\footnote{This is just a name that has nothing to do with
a gravitational mass.} ${\cal M}^k$ and stiffness matrix ${\cal S}^k$ are
given by
\begin{equation}
{\cal M}_{ij}^k = \lprod{\ell_i^k}{\ell_j^k}{k}, \mbox{\hspace*{2em}}
{\cal S}_{ij}^k = \lprod{\ell_i^k}{\frac{d \ell_j^k}{d x}}{k}.
\end{equation}
Multiplying through with the inverse of the mass matrix $({\cal M}^k)^{-1}$
and defining the differentiation matrix ${\cal D}_r^k = ({\cal M}^k)^{-1} {\cal S}^k$
we find
\begin{equation}
\frac{d}{dt}\bm{u}_h^k = -a {\cal D}_r^k \bm{u}_h^k+\left ({\cal M}^k\right )^{-1} \left [ \bm{\ell}^k(x) (a u_h^k-(a u_h)^*)
\right ]_{x_r^k}^{x_l^k}.
\end{equation}
If we now form the 2-element boundary vector $\bm{v}=[v_1, v_2]^T$ where
$v_1 = (a u_h^k -(a u_h)^*)|_{x_r^k}$ and
$v_2 = (a u_h^k -(a u_h)^*)|_{x_l^k}$ (i.e.\ $\bm{v}$ contains the boundary
terms of the element which are coupled to the neighboring elements through
the numerical Lax-Friedrich flux $(a u_h)^*$) and use the properties that
$\bm{\ell}^k(x_r^k) =\delta_{i1}$ and $\bm{\ell}^k(x_l^k)=\delta_{iN_p}$ to
introduce the $N_p\times 2$ vector
\begin{equation}
\bm{b} = \left (
  \begin{array}{cc}
    1 & 0 \\
    0 & 0 \\
    \vdots & \vdots \\
    0 & 1
  \end{array}
\right )
\end{equation}
we can rewrite the boundary term and obtain
\begin{equation}
\frac{d}{dt}\bm{u}_h^k = -a {\cal D}_r^k \bm{u}_h^k+{\cal L}^k\bm{v}
\end{equation}
where the lift matrix ${\cal L}^k$ is defined as
\begin{equation}
{\cal L}^{k} = ({\cal M}^k)^{-1} \bm{b}.
\end{equation}
Given the differentiation matrix and the lift matrix, the RHS evaluation in
the DG scheme consists of construction of the boundary matrix (using e.g.\
the Lax-Friedrich flux) and then performing matrix-vector multiplications
to calculate the spatial derivative and the boundary correction.

The above discussion provides a brief summary of the general idea behind the DG method.  There are
still several technical details that needs to be addressed before it
can be turned into a working numerical algorithm.  For example, we have not
described how the nodal points inside the elements are chosen (there is
a reason the nodes are unequally spaced) and we have left
out any discussion of how to numerically construct the differentiation and
lift matrix. For further details, we refer the reader to \cite{Hesthaven:2007:NDG:1557392}.

\section*{References}
%

\end{document}